\documentclass[10pt,journal,twocolumn]{IEEEtran}
\newif\ifCLASSOPTIONromanappendices \CLASSOPTIONromanappendicestrue
\usepackage{a0size}
\usepackage{amssymb}
\usepackage{multicol}
\usepackage[english]{babel}
\usepackage{epsfig}
\usepackage{bm}
\usepackage{amsfonts,color,amsthm,amsmath, lscape,graphics}
\usepackage{url}
\usepackage{algorithm}
\usepackage{algpseudocode}

\usepackage[font=footnotesize]{caption}
\usepackage{subcaption}
\usepackage{epstopdf}
\usepackage{algorithm}
\usepackage{algpseudocode}
\usepackage{cite}
\usepackage{relsize}

\DeclareFontFamily{U}{matha}{\hyphenchar\font45}
\DeclareFontShape{U}{matha}{m}{n}{
      <5> <6> <7> <8> <9> <10> gen * matha
      <10.95> matha10 <12> <14.4> <17.28> <20.74> <24.88> matha12
      }{}
\DeclareSymbolFont{matha}{U}{matha}{m}{n}
\DeclareMathSymbol{\odiv}         {2}{matha}{"63}

\usepackage{fix2col}
\DeclareMathOperator*{\Maximize}{maximize}
\DeclareMathOperator*{\Minimize}{minimize}

\renewcommand{\figurename}{Fig.}
\addto\captionsenglish{\renewcommand{\figurename}{Fig.}}

\newcommand{\bb}{\mathbf{b}}

\newcommand{\bh}{\mathbf{h}}
\newcommand{\bg}{\mathbf{g}}
\newcommand{\bv}{\mathbf{v}}
\newcommand{\bV}{\mathbf{V}}
\newcommand{\bu}{\mathbf{u}}
\newcommand{\bU}{\mathbf{U}}
\newcommand{\bI}{\mathbf{I}}
\newcommand{\bw}{\mathbf{w}}
\newcommand{\bx}{\mathbf{x}}

\newcommand{\bs}{\mathbf{s}}

\newcommand{\by}{\mathbf{y}}
\newcommand{\bz}{\mathbf{z}}
\newcommand{\bA}{\mathbf{A}}

\newcommand{\bW}{\mathbf{W}}

\newcommand{\bn}{\mathbf{n}}

\renewcommand{\frac}{\dfrac}

\newcommand{\diag}{{\mbox{diag}}}

\definecolor{myOrange}{rgb}{1,0.5,0}
\definecolor{myGreen}{rgb}{0,0.5,0}

\newcommand{\changeN}[1]{{\color{black}#1}}
\newcommand{\changeF}[1]{{\color{black}#1}}
\newcommand{\changeW}[1]{{\color{black}#1}}

\newcommand{\Elemsquare}{\mathbin{\text{$\vcenter{\hbox{\textcircled{\tiny{$2$}}}}$}}}

\begin{document}
\title{Active Sensing for Communications by Learning}

\author{
{Foad~Sohrabi},~\IEEEmembership{Member,~IEEE,}
        Tao~Jiang,~\IEEEmembership{Graduate Student Member,~IEEE,}\\
	Wei~Cui,~\IEEEmembership{Graduate Student Member,~IEEE,}
        and~Wei~Yu,~\IEEEmembership{Fellow,~IEEE}
\thanks{The authors are with The Edward S.\ Rogers Sr.\ Department of Electrical and Computer Engineering, University of Toronto, Toronto, ON M5S 3G4, Canada (e-mails: \{fsohrabi, tjiang, cuiwei2, weiyu\}@ece.utoronto.ca). This work is supported by Huawei Technologies Canada and by the Natural Sciences and Engineering Research Council (NSERC). The source code for this paper is available at: \protect\url{https://github.com/foadsohrabi/DL-ActiveSensing}.
}
}
\maketitle
\begin{abstract}
\changeN{
This paper proposes a deep learning approach to a class of active sensing
problems in wireless communications in which an agent sequentially interacts with an environment over a predetermined number of time frames to gather information in order to perform a sensing or actuation task for maximizing some utility function. In such an active learning setting, the agent needs to design an adaptive sensing strategy sequentially based on the observations made so far. To tackle such a challenging problem in which the dimension of historical observations increases over time, we propose to use a long short-term memory (LSTM) network to exploit the temporal correlations in the sequence of observations and to map each observation to a fixed-size state information vector. We then use a deep neural network (DNN) to map the LSTM state at each time frame to the design of the next measurement step. Finally, we employ another DNN to map the final LSTM state to the desired solution. We investigate the performance of the proposed framework for adaptive channel sensing problems in wireless communications. In particular, we consider the adaptive beamforming problem for mmWave beam alignment and the adaptive reconfigurable intelligent surface sensing problem for reflection alignment. Numerical results demonstrate that the proposed deep active sensing strategy outperforms the existing adaptive or nonadaptive sensing schemes.
}
\end{abstract}


\section{Introduction}
The active sequential learning problem naturally arises in many complex
inference, sensing, and control settings, e.g., tree-search \cite{cameron},
sequential design of experiments \cite{herman} (an example of which is the game of
twenty questions \cite{huanghu}), and the multi-armed bandit \cite{katehakis}
problems.  These problems typically involve an estimation or adaptive control
task which is based on sequential sensing of the environment---in the sense that
instead of passively collecting all the observations of the environment at once,
the agent seeks to \emph{actively} and \emph{sequentially} query the
environment to build up its knowledge about the problem instance, in order to
produce a final optimal solution. In other words, at each time step of the
query procedure, the agent can adaptively design its sensing strategy based on
the information obtained from all the previous steps. The design of active sensing strategies is a highly nontrivial task. The goal of this paper is to develop a unified deep learning framework to design active sensing procedures in a data-driven way.

More specifically, this paper considers an active sensing process in which the
agent seeks to find a solution to an optimization problem of interest by
acquiring sequential measurements of the environment across multiple time
frames. The measurement obtained by the agent in each time frame is a function
of the sensing strategy employed by the agent. To achieve
the best final performance, the agent can adaptively design the sensing
strategy in each time frame based on the available historical observations
obtained in the previous time frames.  Finally, after a fixed number of
measurement acquisition steps, the agent utilizes all the observations to
compute an optimized solution to an estimation or control task. 

An important application of active learning is in channel estimation for
wireless communications. In the initial channel sensing phase for many wireless
protocols, the base station (BS) needs to obtain channel state information
(CSI) for the subsequent communication phase, but the wireless channels are
typically high dimensional and often the BS can only make a low-dimensional
observation based on probing the channel in a specific direction. This paper
aims to show that adopting a data-driven active channel sensing strategy can
significantly improve the system performance, as compared to the conventional
passive approach in which the sensing directions are fixed beforehand. 

In particular, this paper investigates the application of the proposed deep
active learning framework to two wireless channel sensing problems: i)
adaptive beamforming for mmWave beam alignment in a radio-frequency (RF) chain
limited system; and ii) adaptive sensing for reflection alignment in a
reconfigurable intelligent surface (RIS) assisted system. \changeF{Both
applications} involve the optimal sequential sensing of a high-dimensional
analog channel in the low-dimensional digital domain.

\subsection{Main Contributions}
The key challenge in tackling the active learning problem in which the agent
sequentially interacts with an environment is how to summarize the historical
observations over time. In particular, since the dimension of the historical data
(if not processed further) grows over time and the observations are temporally correlated due to
the active sensing operation, constructing \changeF{a sufficient statistic} of the
historical observations (which is crucial for designing the next sensing
strategy and for finding the solution at the final time frame) can be difficult
to do analytically, and \changeF{even if theoretically possible}, can be
computationally costly. In addition, finding an optimal mapping from the
historical observations to the sensing strategy in the next step is a highly
nontrivial task.

This paper proposes a deep learning framework to tackle these challenges.
Specifically, we use a recurrent neural network (RNN) \cite{rnn} to map
the historical observations to a fixed-size representation, called
\textit{state information}, while accounting for the temporal correlation in the sequence of observations. We choose RNN for its capacity to model and summarize sequential observations. As shown in \cite{anton}, RNN is \changeW{a universal} function approximator for any \emph{open dynamic system} including the active learning setting considered in this paper. Moreover, we select the long short-term memory (LSTM) variant of RNN \cite{lstm}, due to its robustness against gradient vanishing and exploding issues \cite{rnn_issue} even under a
relatively large number of time frames. 
We use the hidden state of the LSTM as the state information. 
A deep neural network (DNN) is 
used to map the LSTM hidden state at each time frame to the design of
the sensing action used in the next measurement phase. At the end, 
another DNN is employed to compute the optimized solution
based on the final state of the LSTM cell.

We investigate the application of the proposed framework to two types of
objectives. First, we consider a setup in which the agent is interested in
estimating a parameter of the environment, which is assumed to be static across
the different stages of the active learning process. Assuming that we have
access to the training set including the environment parameter, this first
type of objective can be treated as a supervised learning problem. Second, this
paper also considers a scenario in which the agent is not directly interested
in estimating the static parameter of the environment; instead, the agent aims
to design some control strategy (e.g., beamforming) in order to maximize a
system utility, which is a function of the environment parameter.  In this case,
we assume that we do not have access to the labeled data for the control strategy, and
the proposed deep learning approach needs to be trained in an unsupervised
fashion. We show that the proposed architecture and the training procedure
apply equally well to both types of objectives. 

We now describe two applications of the proposed adaptive sensing strategy to
the high-dimensional analog channel sensing problems in wireless communications.

\subsubsection{Adaptive Beamforming for mmWave Initial Beam Alignment} 

The first application is the directional beam alignment problem for a
single-user mmWave system in which the BS is equipped with massive \changeF{number of} antennas but
only has a single RF chain. In the beam alignment procedure, the BS aims to
either estimate channel parameters such as the angles-of-arrival (AoAs) or to
design a downlink precoder for the subsequent data transmission phase.  
To do so, we use an uplink pilot phase in which the RF-chain-limited BS
observes the user pilots transmitted through the channel via some analog sensing
vectors over multiple time frames. The idea is that we can optimize the sensing
performance by designing the sensing vector sequentially over the time frames as 
a function of the available observations so far. Finding the optimal active 
sensing strategy for the initial access phase, in general, is quite a
challenging task. To make the problem more tractable, most existing methods 
advocate selecting the analog sensing
vectors from a predesigned set of sensing vectors called beamforming codebook
\cite{alkhateeb2014channel,Zhang2017TCOM,Love2017multires,Tara2019Active,Tara2019sequential,Akdim2020spawc}. It has been recently shown that deep learning can be used to
significantly improve the system performance by designing a codebook-free
adaptive sensing strategy \cite{Foad2021JSAC-ML}. However, since the deep
learning framework in \cite{Foad2021JSAC-ML} utilizes the (approximated)
posterior distribution as the state, the application of this method is limited
to the single-path channel model in which the update of the posterior
distribution after observing a new measurement is \changeW{computationally} tractable using the Bayes's rule. 

Unlike the deep learning approach in \cite{Foad2021JSAC-ML}, we show that the
proposed LSTM-based deep active sensing architecture can be used to construct
effective description of the state for a more general multi-path scenario under
different objectives such as AoA estimation and downlink beamforming gain
maximization.  Numerical results show that, even for the AoA estimation problem
in the single-path environment, the proposed approach already achieves better
performance as compared to the algorithm in \cite{Foad2021JSAC-ML}, due to the
fact that the posterior computation in \cite{Foad2021JSAC-ML} has to involve
approximations in order to deal with the unknown fading coefficient. 

A further advantage of the new approach is that it can handle even noncoherent
measurements.  To illustrate this, we apply the proposed approach to the AoA
estimation problem for a noncoherent sensing scenario where the received pilots
are corrupted with phase noise \cite{zhang2019side} and numerically show that
the proposed approach can achieve excellent performance. 

We \changeW{note that a very recent work \cite{dehkordi2021} also} uses 
LSTM-based RNN in mmWave channel estimation, but for beam tracking, which is
different from active sensing. \changeN{In particular, \cite{dehkordi2021}
considers the AoA tracking problem in a line-of-sight environment in which a user moves around, so its AoA changes over time. To track the AoA,
\cite{dehkordi2021} proposes an LSTM-based RNN that captures the temporal
correlation between the AoAs over \changeF{different time frames}. In contrast, this paper considers a mmWave environment in which the AoAs of the user are fixed, and develops an \changeF{LSTM-based} framework to estimate the AoAs over multiple active sensing stages where each sensing stage consists of a low-dimensional observation of the high dimensional channel.}

\subsubsection{Adaptive RIS Sensing for Reflection Alignment}

As a second application, we consider a channel sensing problem for an
RIS-enabled environment in which the reflection coefficients of an intelligent
surface can be configured in order to maximize some system objective, e.g.,
the beamforming gain.  A key requirement for performing such reflection
alignment is to estimate the channel, which is a difficult task due to the fact
that the RIS typically has many elements so the channel is high dimensional, 
but the overall transceiver chains are of much lower dimension. 

The conventional pilot phase for obtaining CSI is designed nonadaptively. 
For example, \cite{9427148} adopts a scheme with random
phase shifts at the RIS over multiple pilot transmissions. To further improve the
performance, \cite{9382000} proposes to leverage location information to learn
a better design for the RIS reflection coefficients in the channel estimation
phase, but it is not a data-driven approach.
In this paper, we make a case that a data-driven approach can result in
a significantly better design of the reflection coefficients in the uplink channel estimation phase. Moreover, an active learning based approach in which the
reflection coefficients in each stage of the channel sensing phase are 
designed as a function of the state information abstracted from the previous 
observations can further improve the overall RIS system performance.

\subsection{Related Works on Active Learning} 

In the machine learning literature, the terminology of \emph{active learning}
has been mainly used for scenarios in which the model actively selects the
input data and the corresponding labels during the training phase
\cite{simontong,claire,settles,kaiyang,yuedeng,tham}. Such active learning
mechanisms are utilized when the training data or labels are expensive to
obtain, so the agent has to be highly selective when procuring training data in
the learning process. Our notion of active learning differs from this line of
work. In particular, in our framework, the agent has the freedom to acquire
new measurements of the underlying channel using different sensing strategies. 
Further, this active measurement procedure occurs consecutively over a
predetermined number of time frames within our model's computation flow for
every problem instance. The agent essentially needs to successively ``focus'' 
onto a subspace of interest in the underlying environment over multiple time
slots, in order to obtain a solution to the overall optimization problem, e.g.,
maximizing a utility function in a wireless communication network. 

In this regard, the active learning setup considered in this paper is closely
related to \emph{attention mechanism} for deep learning. Attention
mechanism is a process in which the model computes its output through actively
zooming in different regions of the input over multiple steps. It has been
widely used in areas including sequence-to-sequence machine translation
\cite{dzmitry,parikh,vaswani,yoon}, image captioning \cite{kelvin,lunhuang},
and reinforcement learning \cite{mott}. Attention mechanism has also been
explored in wireless communication applications, such as for indoor optical
wireless \changeF{communications} \cite{jiayuan}, where an attention layer is added to an
RNN for processing received symbol sequences, as well as for network traffic
prediction \cite{mingli}, where an attention module is designed to selectively
zoom into the spatial-temporal representations of the wireless traffic. Despite
sharing some similarities, the attention \changeW{mechanism based} methods obtain
measurements by focusing on individual regions of the input, e.g., different
parts of the sentences or groups of pixels in the images. This way of attending
to local information is applicable to inputs with clear spatial
structures and divisible sub-parts, which is not the case for the
wireless communication channels considered in this paper.

Furthermore, the proposed active learning framework shares similarities with
some existing techniques in reinforcement learning
\cite{volodymyr,langford,huanghu} in which the 
agent repetitively measures or queries the stationary environment to gradually
build up its knowledge of the environment.  However, reinforcement learning is
well-suited for problems with infinite horizon (or having an absorbing state in
the underlying Markov chain) and the agent seeks to optimize some cumulative reward, while in this paper, we seek to solve problems where the number of acquisition stages is fixed a
priori, and the objective is to find a solution to an optimization problem at
the final acquisition stage.  

\changeN{Lastly, we mention that reference \cite{jayakumar} tackles a reinforcement
learning problem for a partially observed system by proposing a deep learning
framework through which the agent can construct an \emph{approximate
information state}.  More specifically, in \cite{jayakumar}, the agent utilizes
an RNN architecture to construct an effective information state by summarizing
historical information obtained in the sequential measurement procedure. We
note that \cite{jayakumar} assumes a setting for reinforcement learning in 
which the environment evolves over time, thus is different from 
the problem setting of this paper. Nonetheless, the idea in \cite{jayakumar} of
summarizing the previous observations into an information state to guide future
actions is insightful and fits into our application scenarios; in fact, 
\cite{jayakumar} motivated a part of the design in our framework.} 

\subsection{Paper Organization and Notations}

The remainder of the paper is organized as follows. Section~\ref{sec:sys} describes the general system model and active sensing problem formulation. Section~\ref{sec:DAL} presents the proposed active sensing framework. Section~\ref{sec:hybrid} discusses the application of the proposed active sensing method to adaptive beamforming for mmWave beam alignment. Section~\ref{sec:RIS}  considers the application of the proposed active sensing method to adaptive RIS sensing for reflection alignment. Finally, conclusions are drawn in Section~\ref{sec:conc}.

This paper uses lower-case letters for scalar variables, lower-case bold-face letters for vectors, and upper-case bold-face letters for matrices. The real part and the imaginary part of a complex matrix $\bV$ are respectively given by $\Re(\bV)$ and $\Im(\bV)$. The element-wise absolute \changeF{values} and the element-wise \changeF{phases} of a complex vector $\bv$ are respectively given by $|\bv|$ and $\angle{\bv}$. We use the notations $\odiv$ and $(\cdot)^{\Elemsquare}$ to denote the element-wise division and the element-wise square, respectively. Further, we use the superscripts $(\cdot)^\top$, $(\cdot)^{\sf H}$, and $(\cdot)^\dagger$ to denote the transpose, the Hermitian transpose, and the pseudoinverse of a matrix, respectively.  The identity matrix with appropriate dimensions is denoted by $\mathbf{I}$; the all-ones vector with appropriate dimension is denoted by $\mathbf{1}$; $\mathbb{R}^{m\times n}$ denotes an $m$ by $n$ dimensional real space; $\mathbb{C}^{m\times n}$ denotes an $m$ by $n$ dimensional complex space; $\mathcal{CN}(\mathbf{0},\mathbf{R})$ represents the zero-mean circularly symmetric complex Gaussian \changeW{distribution with a covariance} matrix $\mathbf{R}$. The notations $\operatorname{log}_{2}(\cdot)$, $\operatorname{log}_{10}(\cdot)$, and $\mathbb{E} [\cdot] $ represent the binary logarithm, decimal logarithm, and expectation operators, respectively. 
Further, $\|\cdot\|_2$ indicates the Euclidean norm of a vector.
Finally, the rectified linear unit (ReLU), sigmoid, and hyperbolic tangent activation functions are respectively defined as $\operatorname{relu}(x) \triangleq \max(0,x)$, $\operatorname{sigmoid}(x)\triangleq \tfrac{1}{1+e^{-x}}$, and $\operatorname{tanh}(x)\triangleq \tfrac{e^{x}-e^{-x}}{e^{x}+e^{-x}}$.

\section{System Model and Problem Formulation}
\label{sec:sys}

We begin by considering an active sensing setup where an agent interacts with
an environment over $T$ time frames for the purpose of estimating a
parameter $\boldsymbol{\theta} \in \mathbb{R}^N$. In each
time frame $t$, the agent designs a sensing vector
$\mathbf{w}_t\in \mathbb{R}^M$ and subsequently observes a measurement 
$\by_t \in \mathbb{R}^D$, which is a function of the sensing vector
$\bw_t$, the underlying system parameter $\boldsymbol{\theta}$, and possibly 
additional stochastic parameters of the system $\bu_t \in \mathbb{R}^U$ as:
\begin{equation}
\label{eq:input-output}
\by_t = \mathcal{H}\left(\bw_t,\boldsymbol{\theta} , \bu_t \right),  ~~ t=1,\ldots,T,
\end{equation}
where $\mathcal{H}: \mathbb{R}^M \times \mathbb{R}^N \times \mathbb{R}^U \rightarrow \mathbb{R}^D$ determines the input-output relationship of the sensing operation. Here, we assume that the system parameter $\boldsymbol{\theta}$ has a prior distribution and is unknown but fixed over $t$, while the stochastic parameter $\bu_t$ is random and unknown. 
An example of a sensing vector is an analog receive beamformer, which gives rise to a low-dimensional digital measurement.

The agent can sequentially design the sensing vector $\bw_t$ at each time frame $t=1,\ldots,T$, possibly in an adaptive manner. This means that the sensing vector in time frame $t+1$ can be designed according to the historical observations, i.e., the measurements and the sensing vectors prior to time frame $t+1$, as:
\begin{equation}
\bw_{t+1} = \mathcal{G}_t\left(\by_{1:t},\bw_{1:t}\right), ~~ t= 0,\ldots,T-1,
\end{equation}
where $\mathcal{G}_t:  \mathbb{R}^{tD} \times  \mathbb{R}^{tM} \rightarrow \mathbb{R}^M$ is the adaptive sensing strategy adopted by the agent in time frame $t$. Note that since no prior observation is available at time frame $t=0$, we let $\bw_1 = \mathcal{G}_0(\emptyset,\emptyset)$, where $\emptyset$ denotes that $\mathcal{G}_0(\cdot,\cdot)$ accepts no inputs and it always outputs the same initial vector $\bw_1$.

The final estimate of the vector parameter $\boldsymbol{\theta}$ can be obtained as a function of the sensing vectors used in all $T$ time frames and the corresponding observed measurements as: 
\begin{equation}
{\boldsymbol{\hat \theta}} = \mathcal{F}\left(\by_{1:T},\bw_{1:T}\right),
\end{equation}
where $\mathcal{F}: \mathbb{R}^{TD} \times \mathbb{R}^{TM} \rightarrow \mathbb{R}^N$ is the parameter estimation scheme. 
Assuming a distortion measure $\mathcal{E}(\cdot,\cdot)$, the parameter estimation problem can be formulated as:
\begin{subequations}
\label{eq:problem_formulation}
\begin{align}
	\Minimize_{\left\{\mathcal{G}_t(\cdot,\cdot)\right\}_{t=0}^{T-1},\hspace{1pt} \mathcal{F}(\cdot,\cdot) }& \mathbb{E}\left[ \mathcal{E}({\boldsymbol{\hat \theta}} , \boldsymbol{\theta} ) \right]\\
\text{subject to}\hspace{14pt} &\bw_{t+1} = \mathcal{G}_t\left(\by_{1:t},\bw_{1:t}\right),~ t=0,\ldots,T-1,\\
& {\boldsymbol{\hat \theta}} = \mathcal{F}\left(\by_{1:T},\bw_{1:T}\right),
\end{align}
\end{subequations}
where the expectation is over the distribution of all stochastic parameters of the system, i.e., $\boldsymbol{\theta}$ and $\{\bu_t\}_{t=1}^T$.

Problem \eqref{eq:problem_formulation} amounts to a joint optimization over the functions $\left\{\mathcal{G}_t(\cdot,\cdot)\right\}_{t=0}^{T-1}$ and $\mathcal{F(\cdot,\cdot)}$ which are respectively the adaptive sensing strategies in different time frames and the final parameter estimation step. 
In general, solving a problem that involves designing functional expressions is quite challenging. To make the problem more tractable, a common practice is to select the sensing vector $\bw_t$ from a predesigned set of vectors called \emph{codebook}, based on some heuristics. For example, in adaptive beamforming for mmWave initial access, most existing methods employ a hierarchical beamforming codebook design \cite{alkhateeb2014channel}. However, it is shown that such codebook-based sensing is quite suboptimal for the initial access problem \cite{Foad2021JSAC-ML}. 
As already shown in \cite{Foad2021JSAC-ML}, a deep learning strategy
can be used to efficiently design a codebook-free sensing strategy by directly designing the mapping functions $\left\{\mathcal{G}_t\right\}_{t=0}^{T-1}$ to optimize problem \eqref{eq:problem_formulation} in a data-driven fashion.

The approach taken in \cite{Foad2021JSAC-ML}, however, is restricted to the single-path model and further relies on a computation
of posterior distribution of $\boldsymbol{\theta}$ in every time step, which is
computationally complex.  The posterior computation is needed because the
dimension of the inputs to the adaptive sensing functions $\mathcal{G}_t(\cdot,\cdot)$ 
increases over time (if not preprocessed) so that even for moderate values of $T$ the design of functions $\mathcal{G}_t(\cdot,\cdot)$
becomes intractable. To address this issue, the conventional approach 
abstracts the historical observations into a lower-dimensional \emph{state variable}
such as the posterior distribution of $\boldsymbol{\theta}$, which is \changeF{a sufficient statistic} for the estimation problem. Based on the posterior, the sensing vector 
at time $t$ and the final estimator at time $T$ can then be derived. 
However, it is not always easy to compute such a posterior distribution. For
example, in order to compute the posterior, the agent needs to know
the input-output model of the system and the distribution of the stochastic
parameters in \eqref{eq:input-output}. In many practical scenarios, such
information may not be available. Further, even if such information is
available, an efficient posterior computation is only possible for a limited
class of system models. Finally, for scenarios where the system parameter
$\boldsymbol{\theta}$ is a continuous variable, the corresponding posterior
distribution is in the form of a probability distribution function which needs
to be discretized and is difficult to \changeW{compute} accurately.

In this paper, we propose to employ a neural network to automatically {\em learn} 
the state variable. More specifically, since the observations are
sequential and correlated in time, we use \changeW{an RNN} based on
the LSTM architecture \changeW{to capture} the temporal correlation across
the measurements. In each time step, the LSTM also generates a hidden state
variable, which is used to design the corresponding sensing vector. 
In the final stage, the parameter is estimated from the final cell state variable.

The proposed architecture is not limited only to the parameter estimation problem. 
In many applications, 
instead of directly estimating the system parameter $\boldsymbol{\theta}$, 
the agent may be interested in designing a control action 
$\mathbf{v} \in \mathbb{R}^V$ in order to
maximize a utility function of $\boldsymbol{\theta}$ and $\mathbf{v}$.
As an example, in the mmWave initial alignment problem, the control action is
the beamforming vector for subsequent data transmission. Likewise, in the
adaptive RIS sensing problem, where we adaptively design the RIS
reflection coefficients in the sensing phase, the ultimate goal is to use the
information gathered to optimize the RIS reflection coefficients for the subsequent data 
transmission phase. For these settings, the problem of interest can be formulated as:
\begin{subequations}
\label{eq:problem_formulation_unsup}
\begin{align}
\Maximize_{\left\{\mathcal{G}_t(\cdot,\cdot)\right\}_{t=0}^{T-1},\hspace{1pt} \mathcal{F}(\cdot,\cdot) }& \mathbb{E}\left[ \mathcal{J}(\boldsymbol{\theta},\bv) \right]\\
\text{subject to}\hspace{14pt} &\bw_{t+1} = \mathcal{G}_t\left(\by_{1:t},\bw_{1:t}\right),~ t=0,\ldots,T-1,\\
& \bv = \mathcal{F}\left(\by_{1:T},\bw_{1:T}\right),
\end{align}
\end{subequations}
where the expectation is over the distribution of all stochastic parameters of the system, i.e., $\boldsymbol{\theta}$ and $\{\bu_t\}_{t=1}^T$.
In this problem, the mapping $\mathcal{F}: \mathbb{R}^{TD} \times \mathbb{R}^{TM} \rightarrow \mathbb{R}^V$ outputs the designed control action $\bv$, possibly subject to some constraints. We show that with minor modifications to the proposed deep learning framework for solving \eqref{eq:problem_formulation}, we can also tackle the problem \eqref{eq:problem_formulation_unsup}. Note that since in solving \eqref{eq:problem_formulation_unsup}, we may not have access to labeled data for the desired output $\bv$, the training of the deep neural network for solving \eqref{eq:problem_formulation_unsup} falls into the unsupervised learning paradigm.

\section{Deep Active Learning Architecture}
\label{sec:DAL}

In this section, we develop a deep learning framework to solve the active sensing problems \eqref{eq:problem_formulation} and \eqref{eq:problem_formulation_unsup}. In problem formulations \eqref{eq:problem_formulation} and \eqref{eq:problem_formulation_unsup}, the agent utilizes the entire historical observations, i.e., $\left\{\by_{1:t},\bw_{1:t} \right\}$, to design the next sensing vector $\bw_{t+1}$. However, as discussed earlier, since the dimension of the historical observations increases as the time index $t$ increases, using the entire history for constructing the sensing vector is not scalable. To address this issue, it is desirable to abstract useful information from the historical observations (for the purpose of sensing vector design) into a fixed-dimensional state information vector denoted by $\bs_t \in \mathbb{R}^S$. 

 \begin{figure}[t]
 \centering
\includegraphics[width=0.45\textwidth]{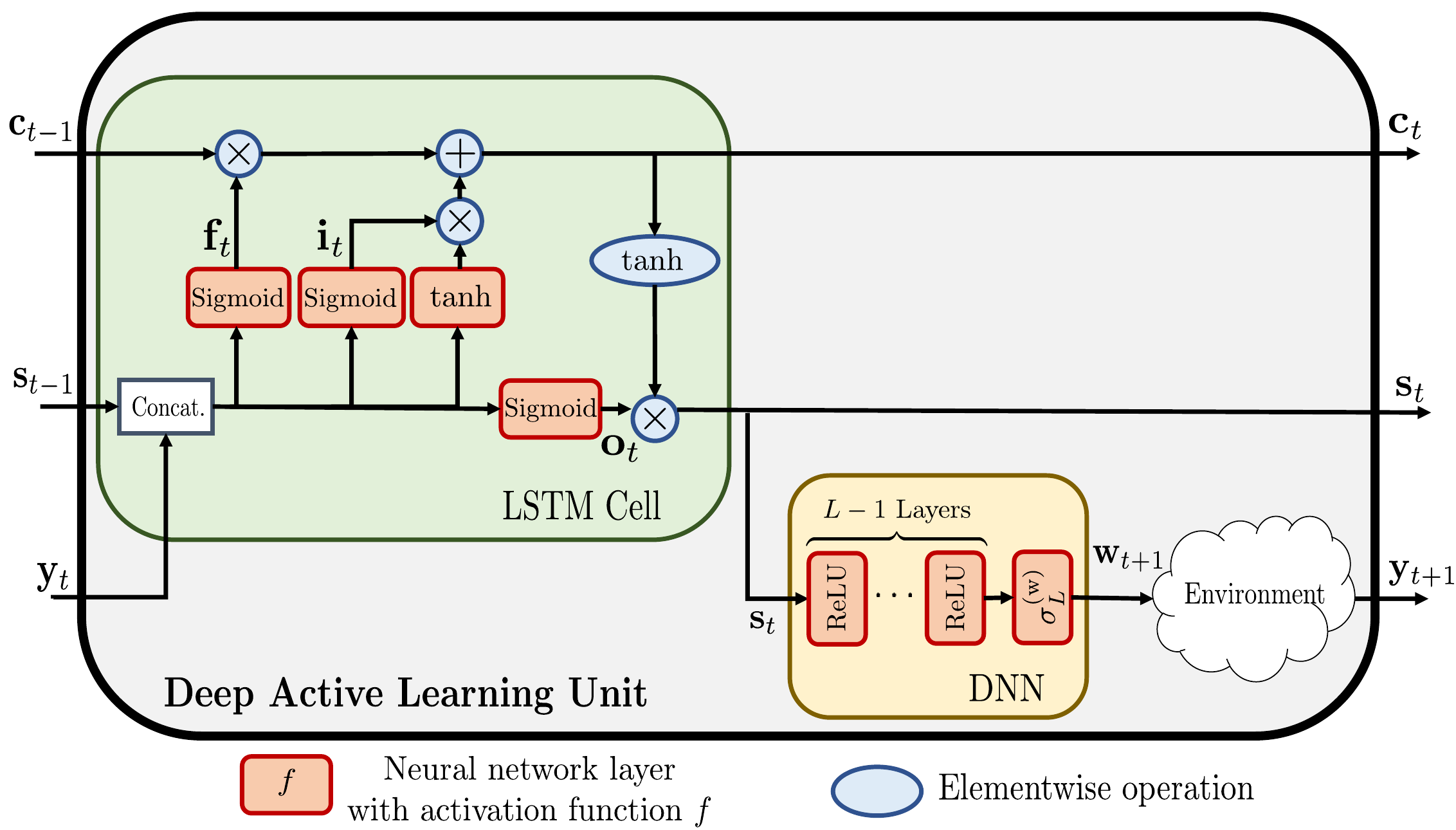} 
\caption{The proposed deep active learning unit for designing the next sensing vector $\bw_{t+1}$ and updating the cell state vector $\mathbf{c}_t$ as well as the hidden state vector $\bs_t$, given the new measurement $\by_{t}$ and the previous state vectors $\mathbf{c}_{t-1}$ and $\mathbf{s}_{t-1}$.}
\label{fig:active_learning_unit}
\end{figure} 

To fix ideas, assuming that the state information $\bs_t$ is given to the agent, we propose to use an $L$-layer \changeW{fully connected} DNN to map the current state $\bs_t$ to the design of the next sensing vector as:
\begin{align} 
{\bw}_{t+1} =& \nonumber\\  \sigma_L^{(\text{w})}& \Big(\bA_L^{(\text{w})}  \sigma_{L-1}^{(\text{w})}\Big(\cdots \sigma_1^{(\text{w})} \Big(\bA_1^{(\text{w})}\bs_t + \bb_1^{(\text{w})} \Big)  \cdots\Big) + \bb_L^{(\text{w})}\Big),
\label{eq:dnn}
\end{align}
where $\left\{\bA_\ell^{(\text{w})}, \bb_\ell^{(\text{w})} \right\}_{\ell=1}^L$ is the set of the trainable weights and biases in the \changeW{fully connected} DNN and $\sigma_\ell^{(\text{w})}$ is the activation function of the $\ell$th layer. Note that the activation functions at the hidden layers, i.e., $\sigma_1^{(\text{w})},\ldots,\sigma_{L-1}^{(\text{w})}$, are design parameters of the network. In this paper, we set the hidden layers' activation functions to the ReLU function, i.e., $\sigma_\ell^{(\text{w})}(\cdot) =\operatorname{relu}(\cdot), ~ \ell = 1,\ldots,L-1$. Further, we can choose an appropriate normalization activation function in the last layer of the DNN to account for any potential constraint that the sensing vector must satisfy.  

The remaining question is how to generate the state information vector $\bs_t$ as a function of the available historical observations at the agent. The notion of \changeF{the} state variable is closely related to the concept of sufficient statistics for estimation. For example, \cite{Tara2019Active} shows that the AoA posterior distribution is \changeF{a sufficient statistic} for the AoA estimation problem in a mmWave environment with a single-dominant path (which we consider in Section~\ref{sec:hybrid}), and proposes to use the posterior distribution as the state information to design the next sensing vector. But as mentioned before, the accurate construction of posterior distribution is quite challenging. This is because to update the posterior distribution after obtaining a new measurement, we need to apply the Bayes's rule which requires the full knowledge of the system model and typically involves multiple integrals which are difficult to evaluate. 

In this paper, we seek to develop a computationally efficient framework that can automatically construct the appropriate state information $\bs_t$ for active sensing problems. To do so, we propose to use an LSTM network which is well-suited to keep track of arbitrary long-term dependencies in the input sequences. In particular, an LSTM cell is employed in each time frame $t$, which takes the new measurement $\by_t$ as the input vector and updates a cell state vector $\mathbf{c}_t$ as well as a hidden state vector $\bs_t$ (which is used as the state information vector in this paper), $\forall t = 0,\ldots,T,$ according to the following equations:
\begin{subequations}
\begin{align}
\mathbf{f}_t &= \operatorname{sigmoid}\left(\mathbf{A}^{(\text{f})} \by_t + \bU^{(\text{f})} \bs_{t-1} + \bb^{(\text{f})}\right),\\
\mathbf{i}_t &= \operatorname{sigmoid}\left(\mathbf{A}^{(\text{i})} \by_t + \bU^{(\text{i})} \bs_{t-1} + \bb^{(\text{i})}\right),\\
\mathbf{o}_t &= \operatorname{sigmoid}\left(\mathbf{A}^{(\text{o})} \by_t + \bU^{(\text{o})} \bs_{t-1} + \bb^{(\text{o})}\right),\\
\mathbf{c}_t &= \mathbf{f}_t \circ \mathbf{c}_{t-1} + \mathbf{i}_t \circ \operatorname{tanh}\left(\mathbf{A}^{(\text{c})} \by_t + \bU^{(\text{c})} \bs_{t-1} + \bb^{(\text{c})}\right),\\ 
\mathbf{s}_t &= \mathbf{o}_t \circ \operatorname{tanh}(\mathbf{c}_t).
	\label{eq:info_state}
\end{align}
\end{subequations}
Here, $\left\{\bA^{(\text{f})},\bA^{(\text{i})},\bA^{(\text{o})},\bA^{(\text{c})},\bU^{(\text{f})},\bU^{(\text{i})},\bU^{(\text{o})},\bU^{(\text{c})}\right\}$ and $\left\{\bb^{(\text{f})},\bb^{(\text{i})},\bb^{(\text{o})},\bb^{(\text{c})}\right\}$ are respectively the trainable weights and biases in the LSTM network. Further, the forget gate's activation vector $\mathbf{f}_t$, input/update gate's activation vector $\mathbf{i}_t$, and output gate's activation vector $\mathbf{o}_t$ are intermediate vectors generated within the LSTM unit to update the cell state vector $\mathbf{c}_t$ and the hidden state vector $\bs_t$. According to the conventional initialization for LSTM networks, we set $\mathbf{c}_{-1} = \mathbf{0}$, $\mathbf{s}_{-1} = \mathbf{0}$, and $\mathbf{y}_0 = \mathbf{1}$.

 \begin{figure*}[t]
 \centering
\includegraphics[width=0.96\textwidth]{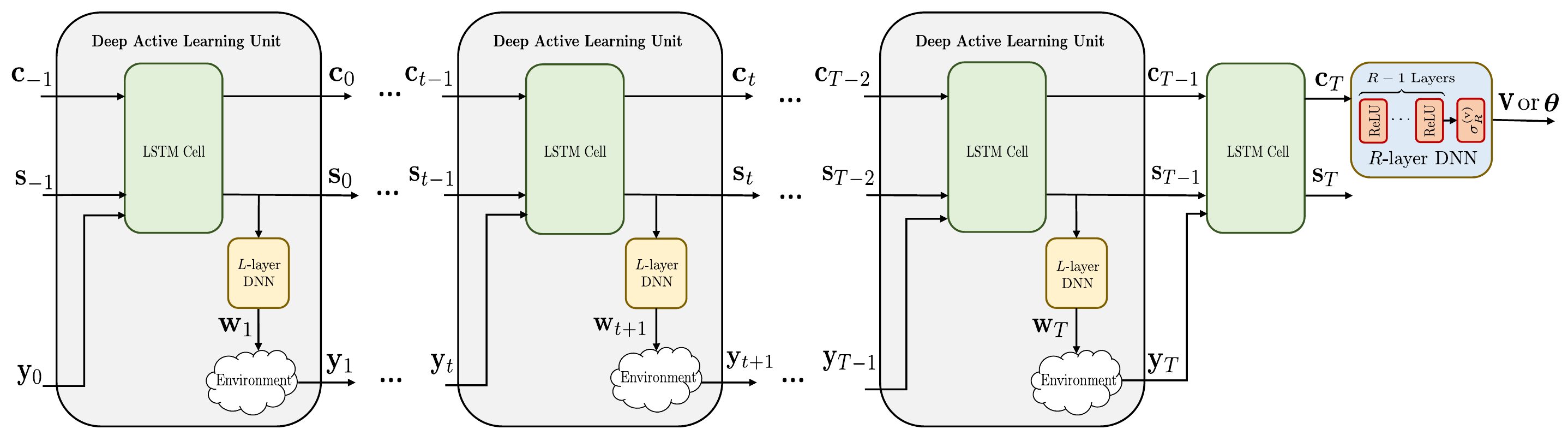}
\caption{The end-to-end architecture of the proposed deep active learning framework.}
\label{fig:active_learning_entire}
\end{figure*} 

In this paper, we propose to use the hidden state vector $\bs_t$ in the LSTM unit constructed according to \eqref{eq:info_state} as the state information passed to the DNN to design the next sensing vector according to \eqref{eq:dnn}. The block diagram of the proposed deep active learning unit at time frame $t$ is illustrated in Fig.~\ref{fig:active_learning_unit}. To learn the trainable parameters of the deep active learning unit in Fig.~\ref{fig:active_learning_unit}, we consider all $T$ sensing stages and regard the end-to-end architecture as a very deep neural network, but with the weights of the LSTM units tied together. 

The overall deep active sensing architecture is as shown in
Fig.~\ref{fig:active_learning_entire}.  The adaptive sensing module implemented
by the DNN as in \eqref{eq:dnn} and the state information constructed by the
LSTM cell as in \eqref{eq:info_state} interact with each other over multiple
stages to enable the design of the sensing vectors, and ultimately either to
estimate the system parameter $\boldsymbol{\theta}$ (i.e., problem
\eqref{eq:problem_formulation}) or 
to maximize the utility function $\mathcal{J}(\boldsymbol{\theta},\bv)$
(i.e., problem \eqref{eq:problem_formulation_unsup}). 
The final estimation or control output is produced by another DNN in time frame $T$ to map the final cell state vector $\mathbf{c}_T$ to the estimate of $\boldsymbol{\theta}$ for problem \eqref{eq:problem_formulation} as:
\begin{align} 
{\boldsymbol{\hat \theta}} =&\nonumber\\  &\sigma_R^{(\text{v})} \Big(\bA_R^{(\text{v})}  \sigma_{R-1}^{(\text{v})}\Big(\cdots \sigma_1^{(\text{v})} \Big(\bA_1^{(\text{v})}\mathbf{c}_T + \bb_1^{(\text{v})} \Big)  \cdots\Big) + \bb_R^{(\text{v})}\Big),
\label{eq:final_estimate}
\end{align}
or to the design of $\mathbf{v}$ for problem \eqref{eq:problem_formulation_unsup} as:
\begin{align} 
{\mathbf{v}} \nonumber =&\\  &\sigma_R^{(\text{v})} \Big(\bA_R^{(\text{v})}  \sigma_{R-1}^{(\text{v})}\Big(\cdots \sigma_1^{(\text{v})} \Big(\bA_1^{(\text{v})}\mathbf{c}_T + \bb_1^{(\text{v})} \Big)  \cdots\Big) + \bb_R^{(\text{v})}\Big),
\label{eq:final_beamformer}
\end{align}
where $\left\{\bA_r^{(\text{v})}, \bb_r^{(\text{v})} \right\}_{\forall r}$ is the set of the trainable weights and biases and ${\sigma}^{(\text{v})}_r$ is the activation function of the $r$th layer. We  set\footnote{\changeN{We employ the $\operatorname{relu}(\cdot)$ activation function due to its computation efficiency and better convergence as compared to other widely used activation functions.}} ${\sigma}^{(\text{v})}_r(\cdot) =\operatorname{relu}(\cdot), ~r = 1,\ldots,{R}-1$ and use the activation function in the last layer to ensure any potential constraint that the final output must satisfy. Finally, we train the neural network architecture in Fig.~\ref{fig:active_learning_entire} by employing \changeW{the stochastic gradient descent algorithm} in order to minimize the empirical average MSE (over the training set) for problem \eqref{eq:problem_formulation} or to maximize average utility function $\mathcal{J}(\cdot,\cdot)$ for problem \eqref{eq:problem_formulation_unsup}.

We emphasize the importance of having a standalone DNN that maps the LSTM state at each time frame to a sensing vector, instead of letting the LSTM directly output the sensing vector. The reasons for this design choice are two-fold. First, from the model capacity perspective, it is too demanding to ask the single-layer LSTM unit to learn both the fixed-size representation of the historical measurements, i.e., the state, as well as the design of the next sensing vector. 
Moreover, the sensing vectors are required to be normalized, for example due to the power constraint in wireless communication settings, but the conventional LSTM unit does not include such normalization. Likewise, we need to employ another DNN at the last stage to map the final state to the estimate of $\boldsymbol{\theta}$ or to the design of $\bv$. This paper uses the final cell state vector $\mathbf{c}_T$ as the input to the DNN at the final stage; see \eqref{eq:final_estimate} and \eqref{eq:final_beamformer}. However, we acknowledge that the final hidden state vector $\mathbf{s}_T$ can alternatively be used.

\changeN{We remark that the architecture proposed in this paper is closely related to the reinforcement learning framework. However, there are some subtle differences that render the direct applications of reinforcement learning to the problem under consideration challenging. First of all, we are looking for a codebook-free sensing design in which the possible action-space size is unbounded, but most of the reinforcement learning literature deals with a setup where the action is chosen from a discrete set of actions, i.e., \changeF{a} codebook. 
Moreover, reinforcement learning is typically suited for problems in which the number of stages is unbounded and the agent aims to optimize a cumulative reward. But for the problems under consideration, it is assumed that the number of time frames that the agent can interact with the environment is fixed and the agent only cares about the final performance which can be exactly computed from the training set. Therefore, the deep learning framework with supervised or unsupervised learning, which is known to have a faster convergence rate and better data efficiency as compared to reinforcement learning, is more appropriate in our setting.}

\section{Active Sensing for mmWave Beam Alignment}
\label{sec:hybrid}
In this section, we consider the adaptive beamforming problem for mmWave initial alignment in an RF-chain-limited system and show that this problem can be formulated in the form of the optimization problems \eqref{eq:problem_formulation} or \eqref{eq:problem_formulation_unsup} for AoA estimation and for downlink beamforming, respectively.

\subsection{System Model and Problem Description}

\begin{figure}[t]
      \centering
     \includegraphics[width=0.45\textwidth]{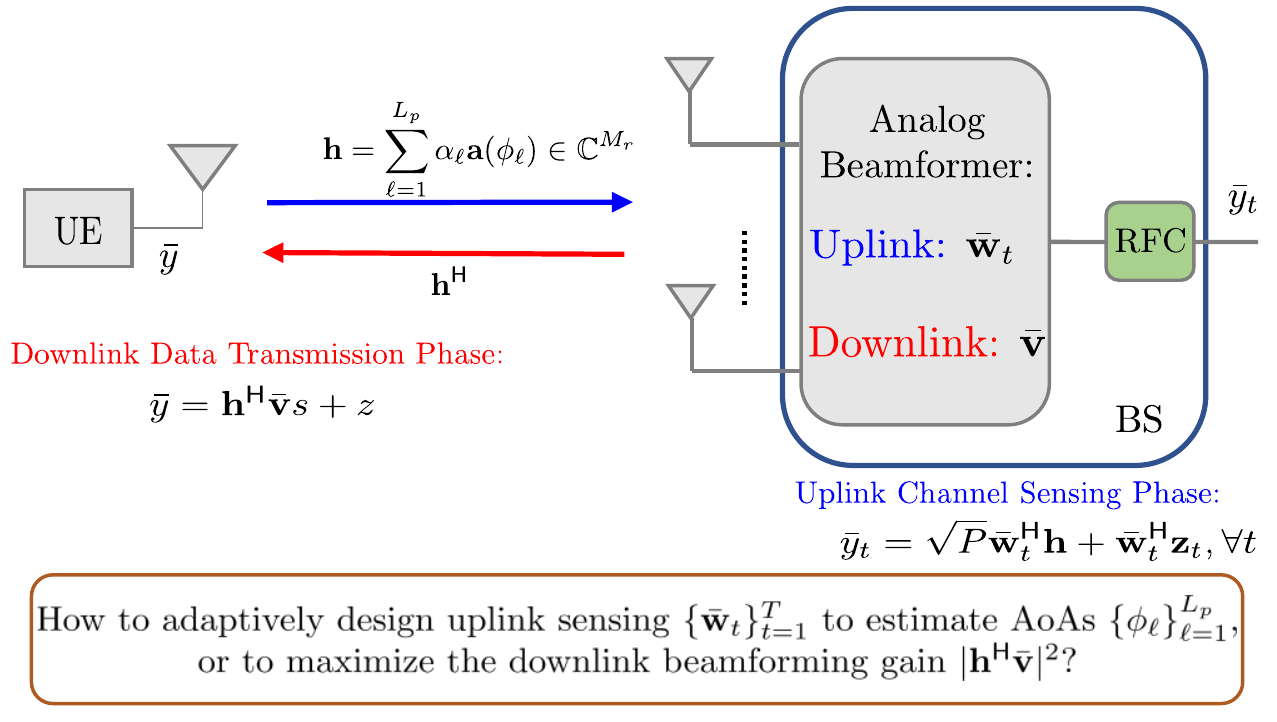} 
     \caption{System model and problem description for adaptive beamforming in the initial access phase of a mmWave communications system.}
     \label{fig:system_hybrid}
\end{figure}

Consider a mmWave communication system in which a BS with $M_r$ antennas and a single RF chain serves a single-antenna user, as illustrated in Fig.~\ref{fig:system_hybrid}. 
To estimate the channel from the user to the BS, 
we use an uplink pilot phase consisting of $T$ time frames, in which the user transmits uplink pilots $\{x_t\}_{t=1}^{T}$ with power $P$, i.e., $x_t = \sqrt{P}, ~\forall t$. 
Due to the single RF chain limitation, the BS observes the received pilot signals only through analog beamforming (or sensing) vectors $\{\bar{\bw}_t\}_{t=1}^{T}$, which can be designed sequentially, possibly as function of the previous observations, i.e.,
\begin{equation}
\label{eq:meausreModel}
\bar{y}_t =\sqrt{P}\hspace{1pt} \bar{\bw}_t^{\sf{H}} \bh + \bar{\bw}_t^{\sf{H}} \bz_t, \quad t=1,\ldots,T,
\end{equation}
where $\bh \in \mathbb{C}^{M_r}$ is the vector of channel gains between the BS and the user, $\bz_t\sim \mathcal{CN}(\mathbf{0},\bI)$ is the additive white Gaussian noise, and without loss of generality we assume that the beamforming vectors \changeF{satisfy} $\|\bar{\bw}_t\|^2_2 = 1, \forall t$. Note that the analog beamformer $\bar{\bw}_t$ in RF-chain-limited systems is typically implemented via a network of phase shifters, and accordingly, the elements of $\bar{\bw}_t$ satisfy the constant modulus constraint, i.e., $|\bar{w}_i^{(t)}| = \tfrac{1}{\sqrt{M_r}}, \forall i,t$, where $\bar{w}_i^{(t)}$ is the $i$th element of $\bar{\bw}_t$. In this paper, we seek to develop a unified deep learning structure that can address both scenarios with and without the constant modulus constraint.

We assume that the mmWave propagation environment between the BS and the user can be modeled geometrically with $L_p$ paths \cite{sohrabi2016hybrid}. Further, by assuming uniform linear array antenna configuration, the channel between the BS and the user can be written as:
\begin{equation}
\bh = \sum_{\ell=1}^{L_p}\alpha_\ell \mathbf{a}(\phi_\ell),
\end{equation}
where $\alpha_\ell \sim \mathcal{CN}(0,1)$ is the complex fading coefficient of the $\ell$th path between the BS and the user, $\phi_\ell \in [\phi_\text{min},\phi_\text{max}]$ is the corresponding AoA, and $\mathbf{a}(\cdot)$ is the array response vector given by
\begin{equation}
\mathbf{a}(\phi_\ell) = \left [ 1, e^{j\tfrac{2\pi }{\lambda} d \sin{\phi_\ell} },..., e^{j(M_r-1)\tfrac{2\pi }{\lambda}d \sin{\phi_\ell} }  \right]^\top,
\end{equation}
where $\lambda$ is the wavelength and $d$ is the antenna spacing.

For a variety of communication applications including fingerprinting-based
localization \cite{Li_localization2008} and downlink beamforming for time
division duplex (TDD) systems with line-of-sight condition \cite{Ng2017}, the
BS is interested in estimating the AoAs, $\{\phi_\ell\}_{\ell=1}^{L_p}$, in
a sensing phase, consisting of $T$ baseband received signals 
$\{\bar{y}_t\}_{t=1}^{T}$.  This sensing phase is sometimes called 
the \textit{initial beam alignment} procedure, in which 
BS can optimize the estimation performance by adaptively designing the sequence
of sensing vectors $\{\bar{\bw}_t\}_{t=1}^{T}$. Specifically, the
beamforming vector in time $t+1$ can be set as a function of the historical
observations including the measurements and the beamforming vectors prior to
time frame $t+1$ as: 
\begin{equation}
\bar{\bw}_{t+1} = \bar{\mathcal{G}}_t\left(\bar{y}_{1:{t}},\bar{\bw}_{1:t} \right), \quad t= 0,\ldots,T-1.
\end{equation}
Here, $\bar{\mathcal{G}}_t : \mathbb{C}^{t}\times \mathbb{C}^{tM_r} \rightarrow \mathbb{C}^{M_r}$ is the adaptive sensing strategy which outputs the next sensing vector that satisfies the required beamforming constraint, i.e., unit power constraint $\|\bar{\bw}_t\|^2_2 = 1, \forall t$, or constant modulus constraint  $|\bar{w}_i^{(t)}| = \tfrac{1}{\sqrt{M_r}}, \forall i,t$.

Further, the estimate for the AoA vector $\boldsymbol{\phi}\triangleq[\phi_1,\ldots,\phi_{L_p}]$ can be obtained as a mapping from the entire available historical observations, i.e., all the sensing vectors and the baseband received signals in $T$ time frames, as:
\begin{equation}
{\boldsymbol{\hat \phi}} = \bar{\mathcal{F}}\left(\bar{y}_{1:T},\bar{\bw}_{1:T} \right),
\end{equation}
where $\bar{\mathcal{F}}: \mathbb{C}^{T}\times \mathbb{C}^{TM_r} \rightarrow [\phi_\text{min},\phi_\text{max}]^{L_p}$ is the AoA estimation scheme. By considering the MSE as the performance metric, the AoA estimation problem for initial beam alignment in a mmWave environment can be written as the following adaptive learning problem:
 \begin{subequations}\label{eq:ex1_sup}
  \begin{align} 
\displaystyle{\Minimize_{\{\bar{\mathcal{G}}_t(\cdot,\cdot)\}_{t=0}^{T-1},\hspace{1pt} \bar{\mathcal{F}}(\cdot,\cdot)}} &~\mathbb{E} \left[ \|| {\boldsymbol{\hat \phi}} -  \boldsymbol{\phi} \|^2_2 \right]\\
\text{subject to}~\hspace{11pt} & ~ \bar{\bw}_{t+1} = \bar{\mathcal{G}}_t\left(\bar{y}_{1:{t}},\bar{\bw}_{1:t} \right), ~  t= 0,\ldots,T-1, \\
	  ~~~~~ &~~ \boldsymbol{\hat{\phi}} = \bar{\mathcal{F}}\left(\bar{y}_{1:T},\bar{\bw}_{1:T} \right),
\end{align}
 \end{subequations}
where the expectation is over the stochastic parameters of the network, i.e., $\boldsymbol{\phi}$, $\boldsymbol{\alpha} \triangleq [\alpha_1,\ldots,\alpha_{L_p}]$, and $\{\bz_t\}_{t=1}^T$.

The AoA estimation problem for initial access in \eqref{eq:ex1_sup} is in the form of the generic adaptive learning problem \eqref{eq:problem_formulation} introduced in Section~\ref{sec:sys}. The only minor difference, which can be easily overcome, is that all parameters and variables in \eqref{eq:problem_formulation} are assumed to be real-valued, while \eqref{eq:ex1_sup} deals with complex signals.
Therefore, the proposed deep learning framework can be used to solve the active sensing problem for AoA estimation in mmWave initial beam alignment. 

We remark that the joint design of the adaptive beamforming strategy and the
AoA estimation scheme by directly solving the problem \eqref{eq:ex1_sup} using
conventional optimization tools is challenging. As mentioned earlier, the
conventional adaptive beamforming schemes consider a codebook-based sensing
strategy, e.g., most of the conventional adaptive schemes for beam alignment
\cite{alkhateeb2014channel,Tara2019Active,Tara2019sequential,Akdim2020spawc}
utilize the hierarchical codebook developed in \cite{alkhateeb2014channel}. The
recent work \cite{Foad2021JSAC-ML} shows that codebook-based beamforming for
initial beam alignment is not optimal, and much more efficient codebook-free
adaptive beamforming can be designed via deep learning. However, the algorithm
in \cite{Foad2021JSAC-ML} is developed based on the simplifying single-path
assumption $L_p=1$, for which the AoA posterior distribution can be calculated
and used as the state variable to design the sensing vectors.
The proposed deep learning approach in this paper can potentially be extended
to a wider class of channel models
with $L_p>1$ for which the computation of the posterior is no longer feasible.

Another advantage of the proposed approach is that it is model free, in
contrast to the model-based approaches (e.g.,
\cite{Tara2019Active,Tara2019sequential,Akdim2020spawc,Foad2021JSAC-ML}).
Thus, the proposed method can be
applied to any type of measurements without the need for the implicit knowledge
of the environment model. For example, the proposed framework can
also be used to tackle the adaptive beamforming for beam alignment with
noncoherent sensing in which the phase information of the signal is not available 
\cite{zhang2019side}. In this case, we can simply consider
the magnitude of the received signal $|\bar{y}_t|$ as the 
input to the deep active learning unit, instead of the complex signal
$\bar{y}_t$.

Finally, we note that for some communication applications, the BS may not be
directly interested in estimating the AoAs. Instead, it seeks to design some
control action such as the downlink beamformer for the subsequent transmission.
This is pertinent in a time-division duplex (TDD) mmWave system with
uplink-downlink channel reciprocity, where the BS can design the downlink
precoder $\bar{\bv}$ based on the uplink CSI learned in the adaptive
sensing phase. If the objective function of the BS is to maximize the
downlink beamforming gain under some precoding constraint (e.g. total power
constraint or constant modulus constraint), the problem can be
written as:
\begin{subequations}
\label{eq:ex1_unsup}
\begin{align}
\Maximize_{\left\{\bar{\mathcal{G}}_t(\cdot,\cdot)\right\}_{t=0}^{T-1},\hspace{1pt} \bar{\mathcal{F}}(\cdot,\cdot) }& \mathbb{E}\left[ \bar{\mathcal{J}}(\bh,\bar{\bv}) \right]\\
\text{subject to}~\hspace{11pt} &  \bar{\bw}_{t+1} = \bar{\mathcal{G}}_t\left(\bar{y}_{1:{t}},\bar{\bw}_{1:t} \right),~ t = 0,\ldots,T-1,\\
	~~ & \bar{\bv} = \bar{\mathcal{F}}\left(\bar{y}_{1:T},\bar{\bw}_{1:T} \right),
\end{align}
\end{subequations}
where 
\begin{equation}
\bar{\mathcal{J}}(\bh,\bar{\bv}) \triangleq 
	|\bh^{\sf H}\bar{\bv}  |^2, 
\end{equation}
and $\bar{\mathcal{F}}: \mathbb{C}^{T} \times \mathbb{C}^{TM_r} \rightarrow \mathbb{C}^{M_r}$ is a mapping \changeW{which} outputs the downlink precoder that satisfies the required beamforming constraint, i.e., unit power constraint $\|\bar{\bv}\|^2_2 = 1$, or constant modulus constraint  $|\bar{v}_i| = \tfrac{1}{\sqrt{M_r}}, \forall i$, where $\bar{v}_i$ is the $i$th element of the downlink precoder $\bar{\mathbf{v}}$. Again, we can see that this problem is in the form of the active learning problem \eqref{eq:problem_formulation_unsup} introduced in Section~\ref{sec:sys}. Therefore, the proposed deep learning framework in Section~\ref{sec:DAL} can be used to tackle the problem \eqref{eq:ex1_unsup}.

\subsection{Implementation Details}
\label{sec:imp_det_hybrid}
In this subsection, we first make an explicit correspondence between input/output of the generic active learning problems, i.e., problems \eqref{eq:problem_formulation} and \eqref{eq:problem_formulation_unsup}, and that of the adaptive beamforming problem for initial beam alignment, i.e., problems \eqref{eq:ex1_sup} and \eqref{eq:ex1_unsup}. For the scenario with coherent measurements, the complex baseband received signal $\bar{y}_t$ is the main component of the input to the proposed deep active learning unit. However, as the existing deep learning libraries only support real-value operations, we need to feed the real representation of $\bar{y}_t$ to the network. Further, in order to employ a common neural network for different values of uplink power $P$, we may choose to also include $P$ (or equivalently signal-to-noise-ratio, i.e., $\operatorname{SNR} \triangleq 10\log_{10}(\tfrac{P}{\sigma^2})$) as the input to the deep active learning unit. 
Numerical results suggest that with this trick we can train one single architecture which achieves a robust performance for different values of SNRs in the AoA estimation problem. However, for the downlink precoding design problem, we observe that training a common architecture that can achieve excellent performance for all SNRs is quite challenging. As a result, we consider the measurement vector at time frame $t$ to be $\by_t := [\Re(\bar{y}_t), \Im(\bar{y}_t), \operatorname{SNR}]^\top$ for the AoA estimation problem and $\by_t:= [\Re(\bar{y}_t), \Im(\bar{y}_t)]^\top$ for the downlink precoding design problem. Similarly, for the scenario with noncoherent measurements, we simply use $\by_t := \left[|\bar{y}_t|, \operatorname{SNR}\right]^\top$ for the AoA estimation problem and $y_t :=|\bar{y}_t|$ for the downlink precoding design problem.

The output of the DNN in the deep active learning unit 
is the real representation of the sensing vector for the next time frame, i.e., $\bw_t := \left[\Re\left(\bar{\bw}_t^\top\right),\Im\left(\bar{\bw}_t^\top\right)\right]^\top$. Moreover, the final output of the proposed deep learning framework is the estimated AoA vector with size $L_p$ for the AoA estimation problem \eqref{eq:ex1_sup}, i.e., ${\boldsymbol{\hat \theta}}:={\boldsymbol{\hat \phi}}$, or the real representation of the beamformer designed for the data transmission phase for the downlink precoding problem \eqref{eq:ex1_unsup}, i.e., $\bv := \left[\Re\left(\bar{\bv}^\top\right),\Im\left(\bar{\bv}^\top\right)\right]^\top$.

In order to ensure that the sensing vectors generated by the deep active learning unit and beamforming vector generated by the final DNN satisfy the required constraints, we need to carefully select the last activation function in the corresponding DNNs, i.e., $\sigma_L^{(\text{w})}(\cdot)$ and $\sigma_R^{(\text{v})}(\cdot)$ in \eqref{eq:dnn}, \eqref{eq:final_estimate}, and \eqref{eq:final_beamformer}. In particular, with only a 2-norm beamforming constraint, we choose the last layer's activation function in the deep active learning unit as:
\begin{equation}
\sigma_L^{(\text{w})}(\underline{\bw}) = \frac{\underline{\bw}}{\| \underline{\bw} \|_2},  ~\forall \underline{\bw} \in \mathbb{R}^{M},
\label{eq:activation_total}
\end{equation}
and for the constant modulus beamforming constraint, we set the last activation function as:
\begin{equation}
\sigma_L^{(\text{w})}(\underline{\bw}) = \sqrt{\tfrac{2}{{M}}}\left[ \underline{\bw}_1^\top \odiv \left|\underline{\bw}_1^{\Elemsquare} + \underline{\bw}_2^{\Elemsquare}\right|, \underline{\bw}_2^\top \odiv \left|\underline{\bw}_1^{\Elemsquare} + \underline{\bw}_2^{\Elemsquare}\right| \right]^\top,
\label{eq:activation_hyb}
\end{equation}
where  $\underline{\bw} = \left[\underline{\bw}_1^\top,\underline{\bw}_2^\top\right]^\top$ with $\underline{\bw}_1,\underline{\bw}_2\in \mathbb{R}^{\tfrac{M}{2}}$. Further, for the downlink precoder design problem in which the output of the final DNN is also a beamforming vector, we set the activation function $\sigma_R^{(\text{v})}(\cdot)$ to be one of the normalization functions in \eqref{eq:activation_total} or \eqref{eq:activation_hyb}, depending on the beamforming constraint. Finally, for the AoA estimation problem where the output is the estimated AoA vector, we employ a linear activation layer for $\sigma_R^{(\text{v})}(\cdot)$.

We implement the proposed deep active learning framework on TensorFlow \cite{tensorflow2016} and Keras \cite{chollet2015} platforms and employ Adam optimizer \cite{adam2014} with a learning rate progressively decreasing from $10^{-3}$ to $10^{-5}$. For the deep active learning unit, we consider an LSTM cell with $S=512$ states and $4$-layer neural networks with dense layers of widths $[1024,1024,1024,M]$ where $M:=2M_r$. In order to map the final cell state to the estimate of $\boldsymbol{\phi}$ for problem \eqref{eq:ex1_sup} (or to the design of downlink precoder $\bar{\bv}$ for problem \eqref{eq:ex1_unsup}), we consider another $2$-layer DNN with widths $[512,D]$ where $D:=L_p$ (or $4$-layer DNN with widths $[1024,1024,1024,D]$ where $D:=2M_r$). The reason for adopting a shallower DNN for the AoA estimation problem as compared to the downlink precoding design problem is that the output dimension is typically much smaller in this case, i.e., $L_p\ll 2M_r$. \changeN{It is worth noting that we consider relatively high-capacity DNNs to investigate the ultimate performance of the proposed framework. However, in our numerical experiments, we also observe that the proposed architecture with much smaller dimensions can still achieve excellent performance, especially in the low-to-medium SNR regime which is the typical scenario for mmWave communication.} For faster convergence, each dense layer is preceded by a batch normalization layer \cite{Ioffe2015}. To investigate the ultimate performance of the proposed approach, we assume that we can generate as many data samples as needed for fully training the proposed deep learning framework. We monitor the performance of the overall neural network architecture during training by evaluating the empirical average of the loss function for an out-of-sample data set of size $10^5$ and keep the network model parameters that have produced the best validation-set performance so far. The training procedure is terminated when the performance of the validation set has not improved over several epochs.

\subsection{Numerical Results}
\label{sec:sims_hyb}

\subsubsection{AoA Estimation}

We now evaluate the performance of the proposed deep active sensing approach for adaptive beamforming for mmWave initial access. We start with the scenario in which the ultimate objective of the beam alignment procedure is to estimate the AoAs. We compare the performance of the proposed method against several existing baselines in the literature. Before presenting the numerical results, we first provide brief explanations of several baseline schemes for AoA estimation. 

\textit{Compressed sensing with fixed beamforming:} In this baseline, the AoA estimation problem is converted to the AoA detection problem by a simplifying (and unrealistic) assumption that the true AoAs are taken from a grid set of size ${N}_\text{CS}$. For this approach, we randomly generate the sensing vectors for all $T$ frames such that they satisfy the required beamforming constraint. If we denote the collection of all $T$ sensing vectors by $\bar{\bW} =  [\bar{\bw}_1,\ldots,\bar{\bw}_T] $ and the collection of response vectors for all ${N}_\text{CS}$ AoA candidates by $\bA_\text{BS} =  [\mathbf{a}(\phi_1), \ldots, \mathbf{a}(\phi_{{N}_\text{CS}})] $, then the baseband received signal at the BS in $T$ time frames can be rewritten as  $\bar{\by} = \bar{\bW}^H \bA_\text{BS} \bx + \bn$, where $\bar{\by}\triangleq [\bar{y}_1,\ldots,\bar{y}_T]^\top$, $\bx$ is an unknown $1$-sparse vector, and $\mathbf{n}$ is the effective Gaussian noise. Accordingly, the AoA detection problem is now equivalent to identifying the support of $\bx$. This sparse recovery problem can be tackled by compressed sensing methods. Here, we adopt a \changeW{widely used} compressed sensing technique called orthogonal matching pursuit (OMP) \cite{Tropp2007OMP} \changeW{as a baseline}. 

\textit{Hierarchical codebook with bisection search (hieBS) \cite{alkhateeb2014channel}:} Under a single-path channel assumption, this on-grid adaptive beamforming method assumes \changeN{that} the true AoA is taken \changeF{from} a grid of size ${N}_\text{hieBS}$ and employs a hierarchical codebook with ${S}_\text{hie}=\log_2({N}_\text{hieBS})$ levels of beam patterns such that each level $s$ consists of a set of 
$2^s$ sensing vectors which partition the AoA search space of $[\phi_\text{min},\phi_\text{max}]$ into $2^s$ sectors. The sensing vector for each sector is designed such that the beamforming gain is almost constant for AoAs within that probing sector, and nearly zero \changeF{elsewhere}. Mathematically speaking, under the 2-norm constraint, the $k$th sensing vector in stage $s$ of the hierarchical codebook is given by
\begin{equation}
\bar{\bw}_{s,k} = \frac{\bA_\text{BS}^\dagger \bg_{s,k}}{\|\bA_\text{BS}^\dagger \bg_{s,k}\|_2},
\label{eq:hie_total_power}
\end{equation}
where $\bg_{s,k}$ is an ${N}_\text{hieBS}$-dimension vector containing $1$'s in the entries $\left\{(k-1)\tfrac{{N}_\text{hieBS}}{2^s}+1,\ldots, k\tfrac{{N}_\text{hieBS}}{2^s}\right\}$, and zeros in the other entries.
For the constant modulus constraint scenario, the modified hierarchical codebook in \cite{Foad2021JSAC-ML} can be employed.
 Finally, the sensing vector in each time frame is selected from a hierarchical codebook using a binary search algorithm, requiring $T = 2{S}_\text{hie} = 2\log_2({N}_\text{hieBS})$ time frames for AoA detection.

\textit{Hierarchical codebook with posterior matching (hiePM) \cite{Tara2019Active}:} 
This method also utilizes the hierarchical codebook in \cite{alkhateeb2014channel}. However, unlike \cite{alkhateeb2014channel}, the proposed method in \cite{Tara2019Active} selects the sensing vectors from the hierarchical beamforming codebook based on the AoA posterior distribution. It should be mentioned that in order to accurately compute the AoA posterior distribution, this approach makes some simplifying assumptions including that there \changeN{exists} only a single path between the BS and the user, the corresponding AoA is taken from a ${N}_{\text{hiePM}}$-point grid, and the fading coefficient value is known at the BS.

\textit{\changeW{Codebook-free posterior based DNN (DNN)} \cite{Foad2021JSAC-ML}:} This approach employs a DNN that maps the posterior distribution (as the state information) at each time frame to the design of the next sensing vector. To deal with the scenario that the fading component is unknown, \cite{Foad2021JSAC-ML} proposes to use an estimate of the fading coefficients (e.g., via Kalman tracking) to compute an approximation of the posterior distribution. However, to form the posterior distribution update rule, \changeW{this method} still requires the single-path channel model assumption. To implement this \changeW{deep learning based} approach, we follow the implementation details provided in \cite{Foad2021JSAC-ML}.

We start by examining the performance of the proposed method and the above baselines in an environment with $L_p=1$ path between the BS and the user, so that the result can be compared with the baselines, as some of the above baselines can be applied only to the single dominant path channel model. 
The AoA corresponding to this single path is uniformly taken from the interval $[\phi_\text{min},\phi_\text{max}] = [-60^\circ, 60^\circ]$, and the number of BS antenna is set to $M_r=64$. In order to have a fair comparison between different methods, we consider that the number of uplink pilot \changeF{transmissions} (i.e., number of time frames in the beam alignment phase) is $T=2{S}_\text{hie}=14$ where ${S}_\text{hie}=7$ is the number \changeF{of} beam pattern levels in the considered hierarchical codebook. For the \changeW{compressed sensing based} method and the hiePM algorithm, which have on-grid AoA assumption, we set the grid size to ${N}_{\text{CS}}={N}_{\text{hiePM}}=2560$.

\begin{figure}[t]
      \centering
      \includegraphics[width=0.5\textwidth]{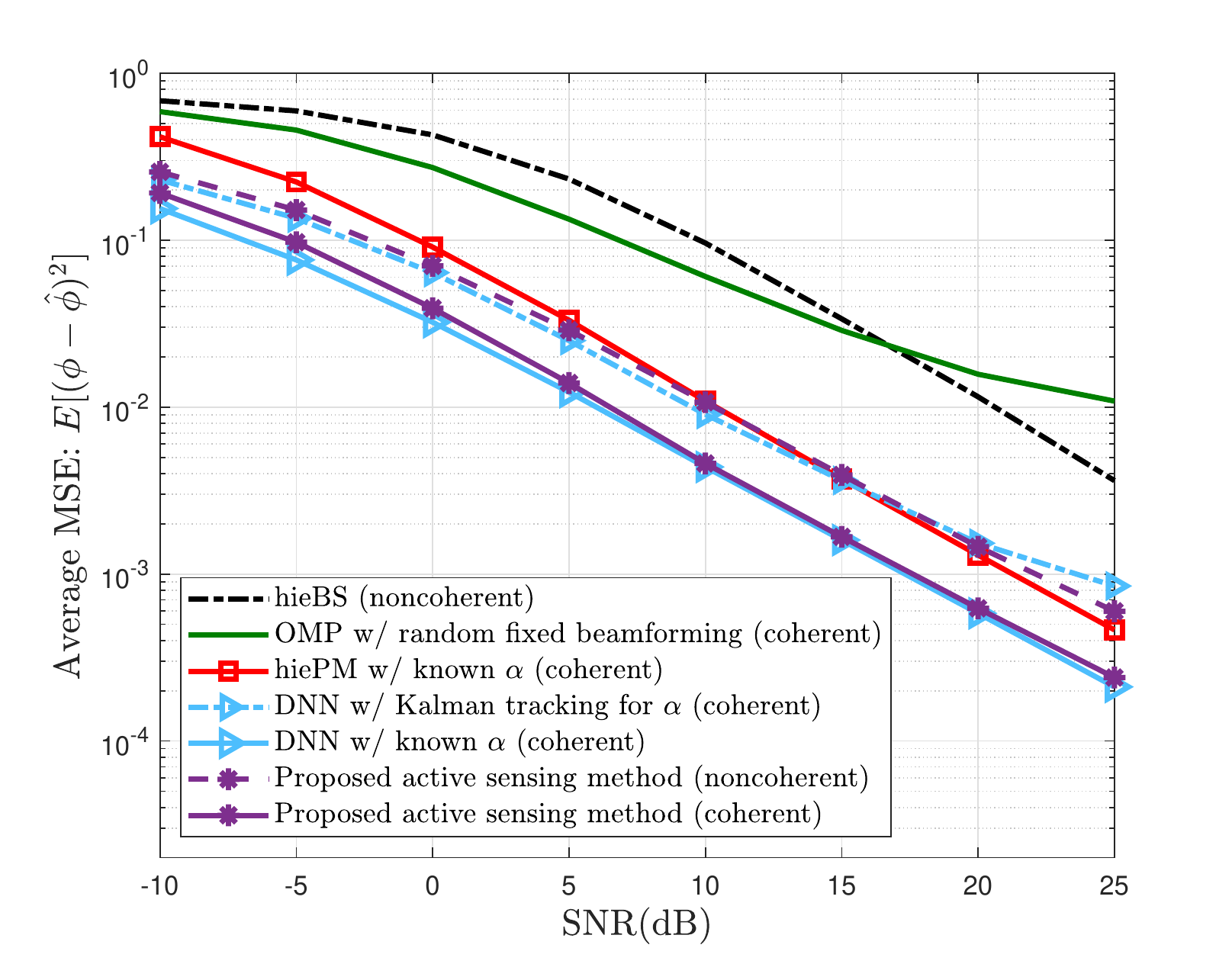} 
      \caption{Average MSE versus SNR for different beam alignment methods in a system with
$M_r = 64$, $L_p=1$, $T = 14$, and $\phi \in [-60^{\circ},60^{\circ}]$. In this experiment, the sensing
vectors must satisfy the 2-norm constraint.}
\label{fig:sim_aoa_est_dig}
\end{figure}

In the first experiment, we consider sensing vectors which need to satisfy only the 2-norm constraint. In Fig.~\ref{fig:sim_aoa_est_dig}, we plot the average MSE for different methods against the SNR. Fig.~\ref{fig:sim_aoa_est_dig} confirms that in general the adaptive sensing strategies can achieve better AoA estimation performance in mmWave beam alignment procedure as compared to the nonadaptive beamforming strategy followed by a compressed sensing channel estimation approach, i.e., OMP. Further, Fig.~\ref{fig:sim_aoa_est_dig} illustrates that the best performance is achieved by the \changeW{DNN} benchmark under the unrealistic assumption that the fading coefficient is perfectly known. Note that when the fading coefficient is known, the posterior distribution, which provides sufficient statistics for AoA estimation, can be accurately computed and used as the state information by the \changeW{DNN} method. 
Interestingly, the proposed active learning method (with coherent sensing but without knowledge of $\alpha$) that learns the state variable using an LSTM approaches the performance of \changeW{DNN} with known $\alpha$. This indicates that the proposed framework is capable of learning the effective state in a data-driven fashion for the adaptive sensing problems for AoA estimation. Finally, as mentioned earlier, the proposed deep active learning framework can also be used for the noncoherent sensing setup. From Fig.~\ref{fig:sim_aoa_est_dig}, we can see that the proposed approach with noncoherent sensing, which has only access to $|\bar{y}_t|$ instead of $\bar{y}_t$, can achieve comparable performance to the state-of-the-art adaptive beamforming strategies with coherent sensing, e.g., \changeW{DNN} with Kalman tracking for $\alpha$.

\begin{figure}[t]
      \centering
      \includegraphics[width=0.5\textwidth]{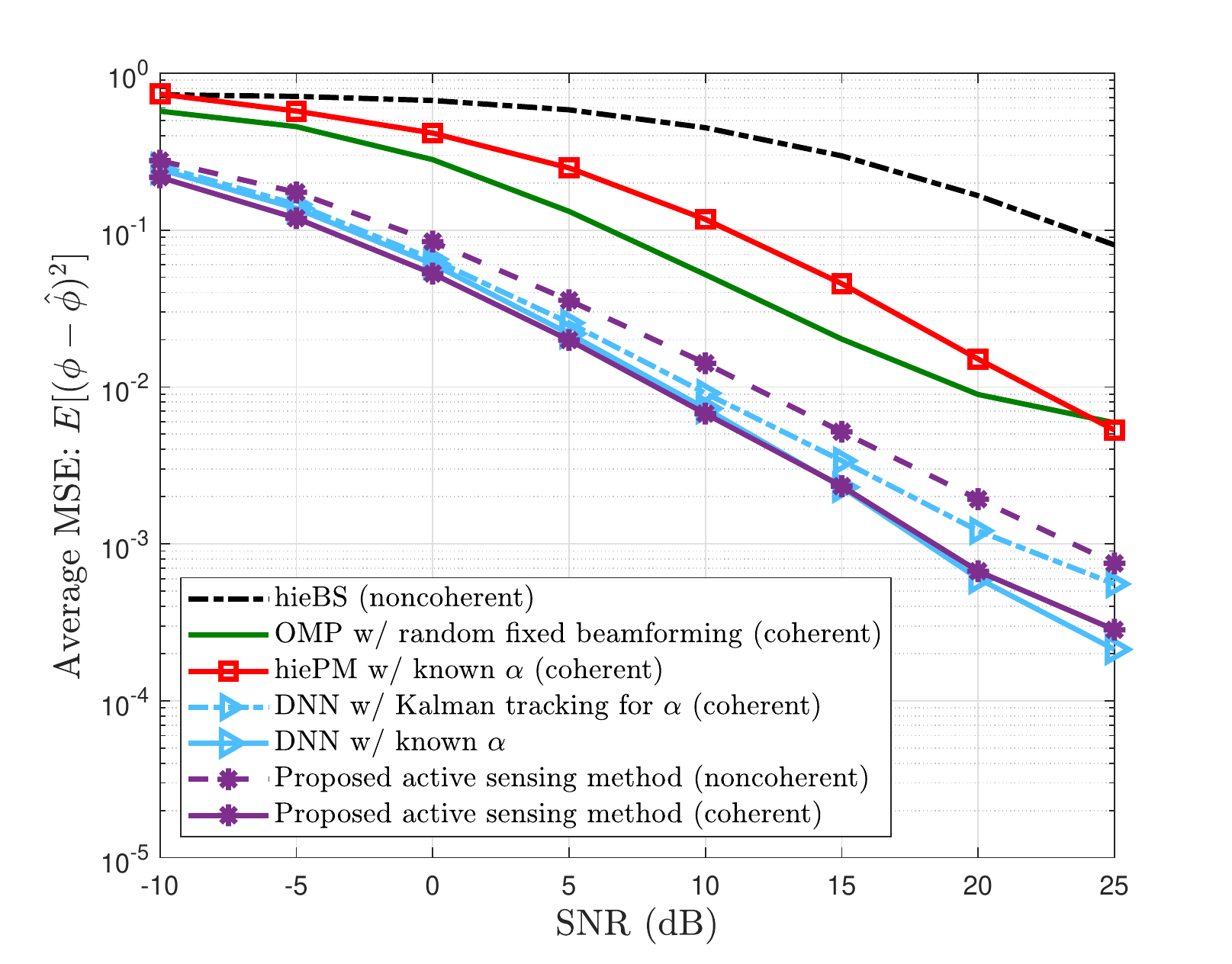} 
      \caption{Average MSE versus SNR for different beam alignment methods in a system with
$M_r = 64$, $L_p=1$, $T = 14$, and $\phi \in [-60^{\circ},60^{\circ}]$. In this experiment, the sensing vectors must satisfy the constant modulus constraint.}
\label{fig:sim_aoa_est_dig_unit_modulus}
\end{figure}

In the next experiment, we consider the scenario that the analog sensing vectors are realized using a network of phase shifters, thus satisfying the constant modulus constraint. Fig.~\ref{fig:sim_aoa_est_dig_unit_modulus} shows that the performance of the methods utilizing the hierarchical codebook, i.e., hiePM and hieBS, is degraded significantly under the constant modulus constraint. However, the \changeW{deep learning based} approaches, i.e., the proposed approach and the \changeW{DNN} method, still exhibit excellent performance under the constant modulus constraint, indicating that such codebook-free frameworks can more efficiently deal with such practical constraints as compared to conventional codebook-based beamforming methods.

\begin{figure}[t]
      \centering
      \includegraphics[width=0.5\textwidth]{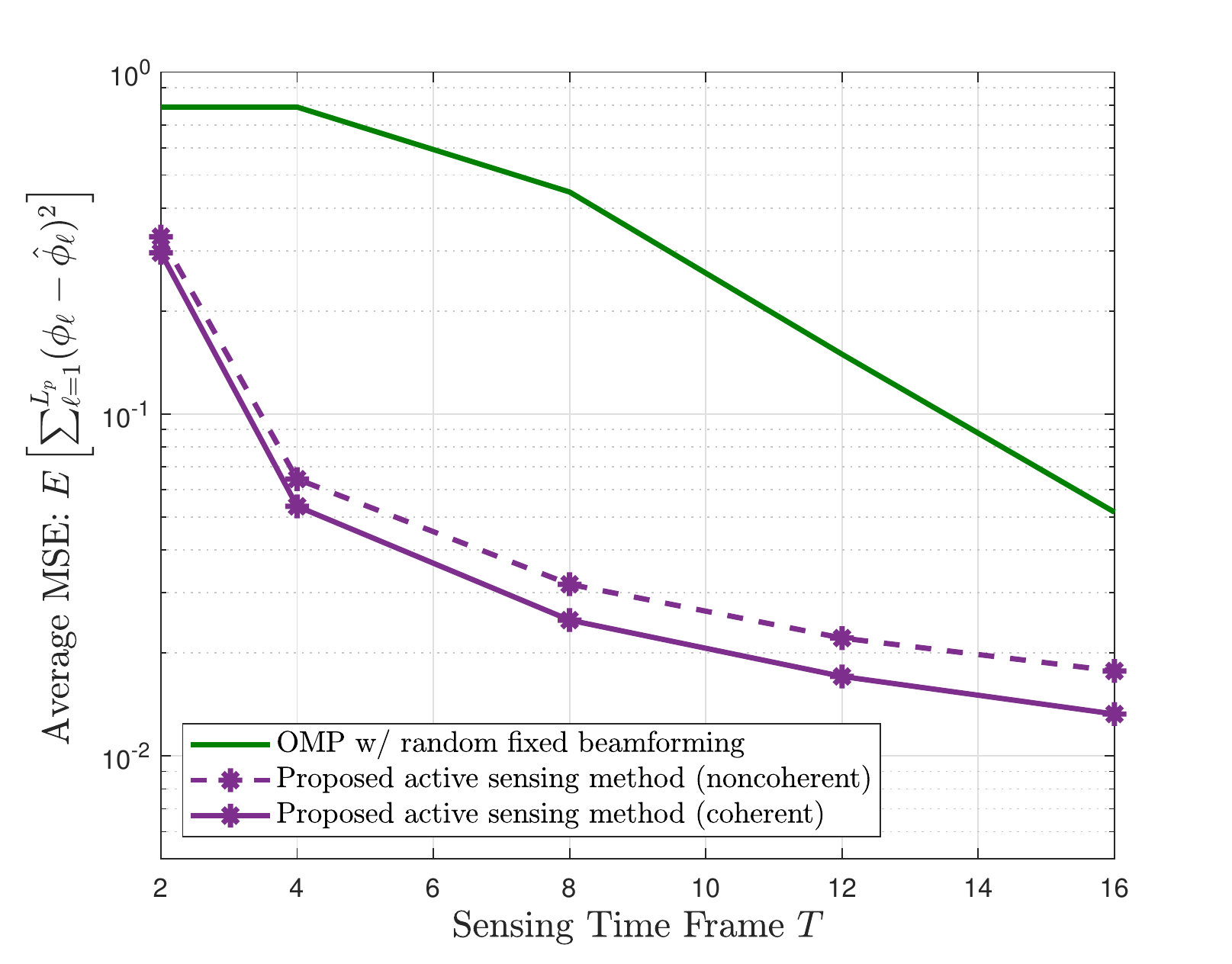} 
      \caption{Average MSE versus sensing time frames $T$ for two AoAs in a system with
$M_r = 64$, $L_p=2$, $\operatorname{SNR}=25$dB, and $\phi_1, \phi_2 \in [-60^{\circ},60^{\circ}]$. In this experiment, the sensing vectors must satisfy the 2-norm constraint.}
\label{fig:sim_aoa_est_dig_unit_modulus_multipath}
\end{figure}

The AoA estimation problem in a multi-path environment where multiple AoAs need
to be estimated is considerably more challenging. 
Fig.~\ref{fig:sim_aoa_est_dig_unit_modulus_multipath} plots the average MSE of
two AoAs in an environment with $L_p=2$ at a fixed SNR of 25dB, but as a function
of the total number of sensing operations. First, we remark that if, in the 
training phase, the true AoAs are sorted according to the strength of their 
corresponding fading coefficients, this can significantly accelerate the 
training process; note that such ordering information is not needed in the
testing phase.  Here, we only compare the performance of the proposed approach
with that of the OMP channel recovery approach with fixed random beamforming
since the other baselines are mainly developed for the single-path channel model.  
By comparing Fig.~\ref{fig:sim_aoa_est_dig} in which $L_p=1$ and
Fig.~\ref{fig:sim_aoa_est_dig_unit_modulus_multipath} in which $L_p=2$ at
comparable SNR and sensing time frames, we see that the performance of all
methods is degraded when we have more AoAs to recover. However, the proposed
approach with either coherent and non-coherent sensing still considerably
outperforms the compressed sensing based AoA estimation with fixed sensing vectors,
especially with short sensing time frames, indicating the benefit of active sensing
for quickly narrowing down the AoAs. 
The improvement in MSE can be an order of magnitude (e.g., at SNR of 10dB); or 
at fixed MSE of $10^{-1}$, the required number of sensing time frames can be reduced
from 13 to 3, which is a dramatic improvement.

We remark that when the number of AoAs increases, the training of the neural network becomes more challenging.

\subsubsection{Downlink Beamformer Design}

In the next experiment, instead of estimating the AoAs, we evaluate the performance of the proposed active learning approach for the beamforming gain maximization problem \eqref{eq:ex1_unsup} in a multi-path channel environment. As comparison, we consider the following baselines.

\textit{Maximum-ratio transmission (MRT) with perfect CSI:} Assuming that the perfect CSI \changeN{(i.e., $\bh$)} is available at the BS, the optimal single-user downlink precoder under the unit power constraint is given by the MRT scheme, i.e.,
$\bar{\bv}^\star = {\bh}/{\|\bh\|_2}$.
This full-CSI case provides a performance upper bound. 

\textit{MRT with compressed sensing CSI estimation:} In this nonadaptive scheme, we use random sensing vectors. Then, we apply a compressed sensing method (e.g., OMP) to the received pilots in order to recover the sparse channel \changeW{with $L_p$ paths}. Finally, the MRT precoder is constructed based on the estimated channel. 

\begin{figure}[t]
      \centering
      \includegraphics[width=0.5\textwidth]{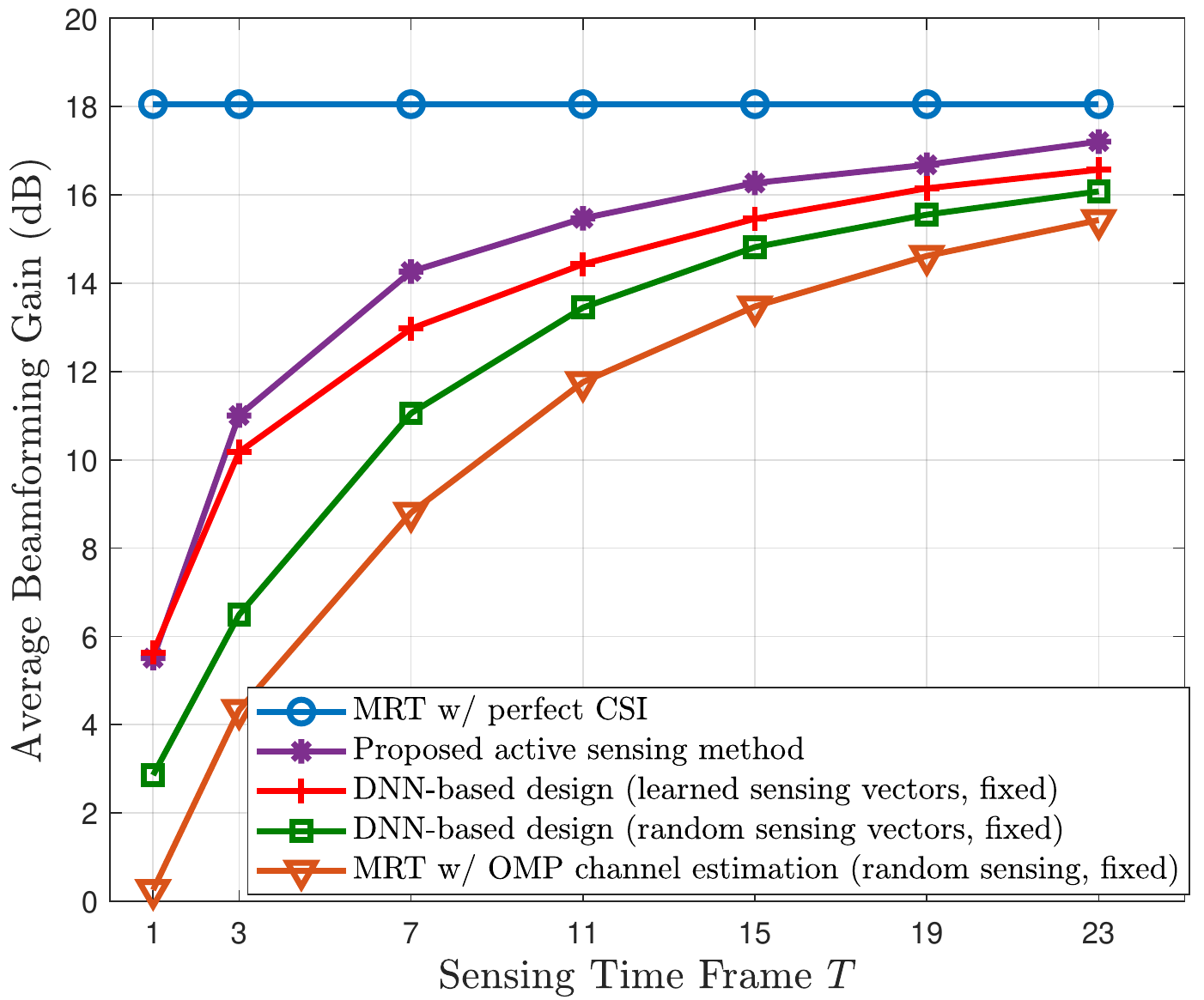} 
      \caption{Average beamforming gain in dB versus sensing time frames $T$ for different methods in a system with
$M_r = 64$, $\operatorname{SNR} = 0$dB, $L_p=3$, and $\phi_1,\phi_2,\phi_3 \in [-60^{\circ},60^{\circ}]$. In this experiment, the sensing
vectors of each method satisfy the 2-norm constraint.}
      \label{fig:sim_multi_AoA}
 \end{figure}

\textit{DNN-based design with nonadaptive sensing \cite{KareemGlobecom2020}:} In this approach, the received pilots observed at the BS in the uplink pilot phase using fixed sensing vectors are directly mapped to the design of the downlink precoder via a \changeW{fully connected} DNN. 
We investigate two variants of this method: (i) the sensing vectors are chosen at random; (ii) the sensing vectors are learned according to the realizations of the channels in the DNN training phase.
To train this DNN, we use the negative of the beamforming gain as the loss function. In the simulations, we implement this approach by employing a 4-layer DNN with widths $[1024,1024,1024,2M_r]$. 

Fig.~\ref{fig:sim_multi_AoA} shows the average beamforming gain in dB versus
the sensing time frames $T$ for different methods in a system with $M_r=64$,
$\operatorname{SNR} = 0$dB, $L_p=3$, and $\phi_1, \phi_2, \phi_3 \in [-60^{\circ},60^{\circ}]$.
From Fig.~\ref{fig:sim_multi_AoA}, it can be seen that the DNN-based methods
that bypass the explicit channel estimation phase are all superior to the
traditional two-step approach of channel estimation followed by downlink
precoding design. Further, Fig.~\ref{fig:sim_multi_AoA} illustrates the gain of
the active sensing strategy in the proposed method as compared to the
DNN-based method with either random fixed sensing vectors or learned sensing
vectors. For example, to achieve the beamforming gain within $2$dB gap from the
optimal beamforming gain in the perfect CSI benchmark, the proposed approach
requires only $15$ time frames, while the other DNN with nonadaptive sensing
needs more than $19$ and $23$ time frames when using random sensing and learned
sensing vectors, respectively. This means that we can save in the overhead of the
uplink pilot phase by adopting the proposed deep active learning method.

To summarize, there is considerable performance gain for the proposed active 
sensing strategy as compared to conventional compressed sensing strategy for 
channel estimation. The performance improvement comes from three sources:
(i) designing sensing vectors for direct optimization of system objective;
(ii) designing sensing vectors in a data-driven fashion according to the 
channel realizations in the training set; (iii) designing sensing vectors 
using active learning techniques. The best performance is obtained when the
sensing vectors \changeW{are both adaptive to the channel environment and sequentially designed using an active learning approach.}

\begin{figure}[t]
      \centering
      \includegraphics[width=0.226\textwidth]{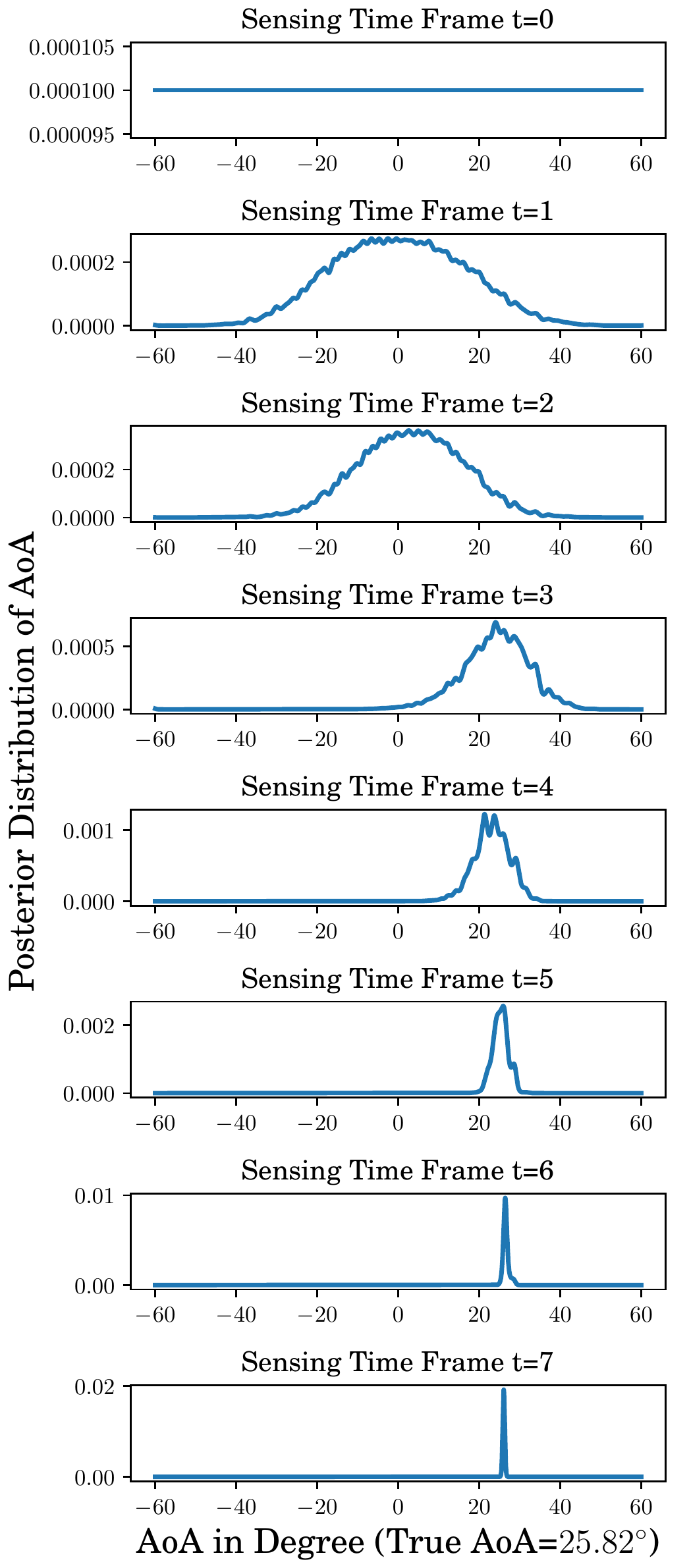} 
      \hspace{0.1cm}
      \includegraphics[width=0.242\textwidth]{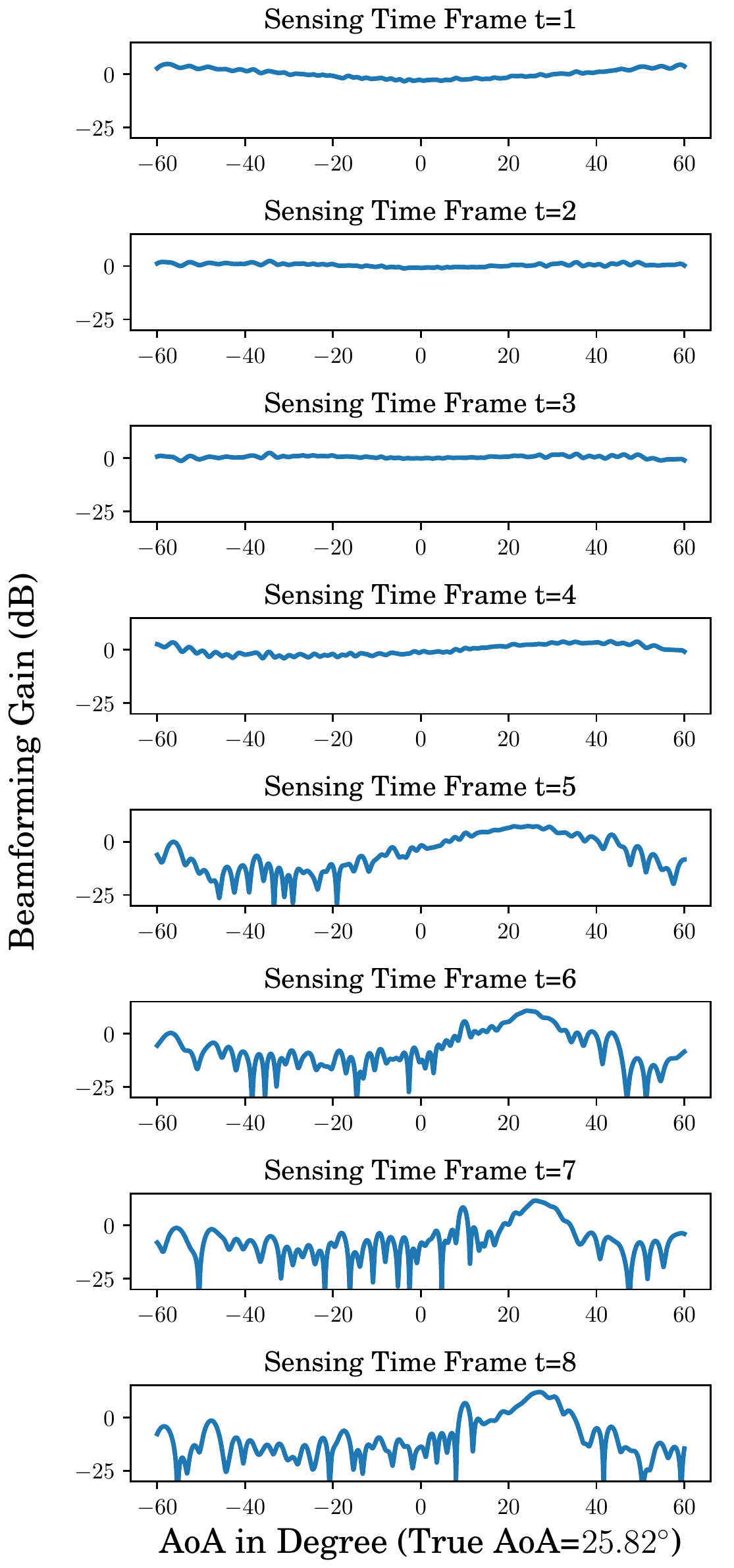} 
      \caption{Posterior distributions of the AoA and the beamforming patterns of the sensing vectors learned from the proposed active sensing framework in a system with $\operatorname{SNR}=0$dB, $M_r = 64$, $L_p=1$, and $T = 12$. In this experiment, the sensing vectors satisfy the 2-norm constraint.}
\label{fig:pos_lstm_tau12}
\end{figure}    

{\color{black}{\subsection{Interpretation of the Learned Active Sensing Strategy}
To visualize the solution learned by the proposed deep active learning framework, in Fig.~\ref{fig:pos_lstm_tau12}, we plot the beamforming pattern as well as the AoA posterior distribution at the end of the $t$-th sensing stage with the sensing vector designed by the proposed deep learning method for estimating the AoA. Specifically,  we consider a single path channel with AoA $\phi = 25.82^{\circ}$ and fading coefficient $\alpha=1$. The system parameters are $\operatorname{SNR}=0$dB, $M_r = 64$, $L_p=1$, and $T = 12$. Further, to compute the AoA posterior distribution, we make a simplifying assumption that the value of the fading coefficient, i.e., $\alpha =1$, is known. From Fig.~\ref{fig:pos_lstm_tau12}, we can see that as the number of sensing stages increases, the posterior distribution gradually converges from a uniform distribution to a highly concentrated distribution with a peak at the true AoA $\phi = 25.82^{\circ}$. Interestingly, the sensing vectors constructed by the active learning framework gradually learn to focus their energy in the direction of the true AoA. This observation suggests that the proposed active learning framework indeed produces meaningful sensing vectors.}}


\section{Adaptive Channel Sensing for Reflection Alignment with RIS}
\label{sec:RIS}

As a second application, we now consider an RIS-assisted system in which
the RIS is tuned adaptively in the channel sensing phase, and the ultimate goal
is to optimize the RIS configuration in the subsequent data transmission phase. 
Again, we show that being adaptive to the channel environment and performing
active sequential estimation both have advantages.

\subsection{System Model and Problem Description}

\begin{figure}[t]
      \centering
      \includegraphics[width=0.49\textwidth]{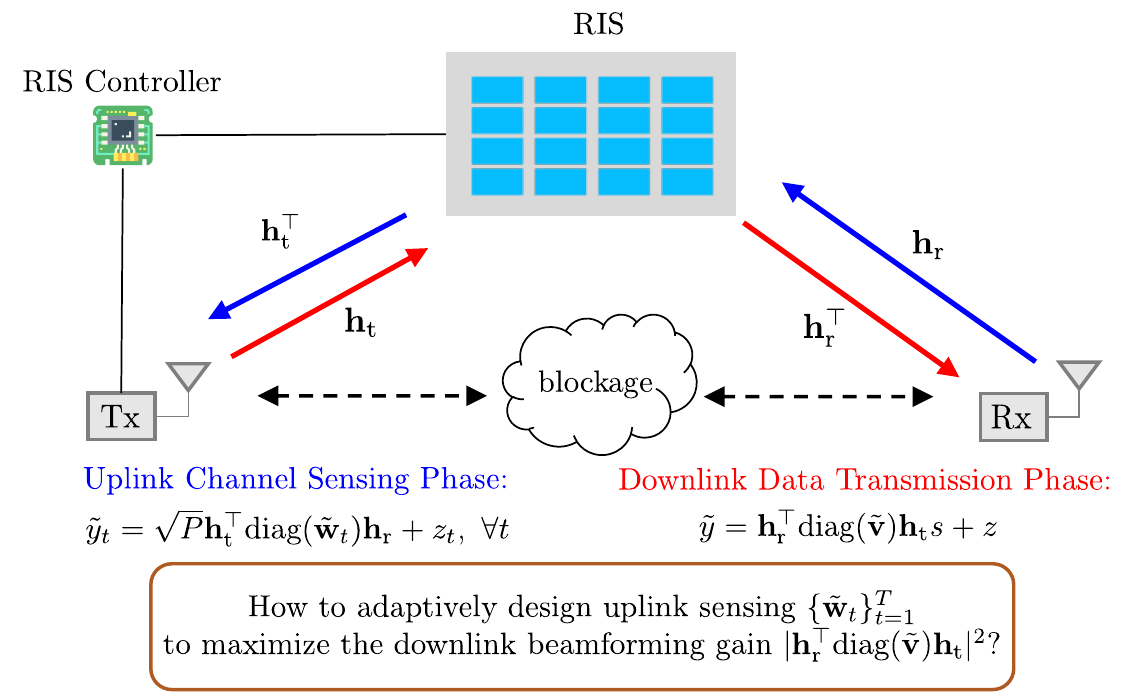} 
      \caption{System model and problem description for adaptive sensing in an RIS-assisted system.}
      \label{fig:system_ris}
 \end{figure}

Consider a point-to-point communication setup in which a single-antenna
transmitter seeks to communicate with a single-antenna receiver, but the
direct channel is blocked, and the communication needs to take place in 
a reflective path with adjustable phase shifts through an RIS. 
The RIS is equipped with $N_r$ passive elements, whose reflective phases can
be configured by a centralized controller, as shown in Fig.~\ref{fig:system_ris}.

Let $\mathbf{h}_{{\rm t}}\in\mathbb{C}^{N_r}$ denote the channel between the transmitter and the RIS, and $\mathbf h_{{\rm r}}\in\mathbb{C}^{N_r}$ denote the channel between the RIS and the receiver. Further, let the vector $\tilde{\mathbf v} \triangleq [e^{j\mathcal{\omega}_1 },\cdots, e^{j\omega_{N_r}}]^\top\in\mathbb{C}^{N_r}$ denote the reflection coefficients of the RIS, where $\omega_i\in [0,2\pi)$ is the phase shift of the $i$th element at the RIS. With the direct channel between the transmitter and receiver blocked, the received signal $\tilde{y}$ in the data transmission phase can be modeled as:
\begin{subequations}
      \begin{align}
            \tilde{y} &=\mathbf h_{{\rm r}}^{\top}\diag(\tilde{\mathbf{v}}) \mathbf h_{{\rm t}} s+z\\
              &=\mathbf h_{\rm c}^{\top} \tilde{\mathbf{v}} s+z,
       \end{align}
\end{subequations}
 where $\mathbf h_{\rm c} \triangleq \diag( \mathbf h_{{\rm t}} ) \mathbf h_{{\rm r}}$ is the cascaded effective channel between the transmitter and the receiver, and $z\sim\mathcal{CN}(0,\sigma^2)$ is the additive Gaussian noise at the receiver. 
For such a system the spectral efficiency (rate) is given by:
\begin{equation}
R = \log_2\left(1+\frac{|\mathbf h_{\rm c}^{\top} \tilde{\mathbf v} |^2}{\sigma^2}\right).
\end{equation}

The transmitter can optimize the spectral efficiency of the system by designing the RIS passive elements $\tilde{\mathbf{v}}$ so as to maximize the downlink beamforming gain, so that the overall objective function is 
\begin{equation}
\tilde{\mathcal{J}}(\tilde{\mathbf{v}}, \mathbf{h}_{\rm c}) \triangleq |\mathbf h^{\top}_{\rm c}\tilde{\mathbf v} |^2.
\end{equation}

In order to design the optimal RIS configuration $\tilde{\mathbf v}$ in the
data transmission phase, it is crucial for the transmitter to have some
knowledge of the effective instantaneous channel $\mathbf h_{\rm c}$, which
must be acquired in a channel estimation stage. 
More specifically, assuming a TDD system with uplink-downlink
channel reciprocity, we use an uplink channel sensing
phase over $T$ time frames prior to the data transmission phase, where the
transmitter observes the pilots sent by the receiver (i.e., $x_t = \sqrt{P}$) in
time frame $t$ as:
\begin{align}
     \tilde{y}_t &=\sqrt{P}\hspace{1pt}\tilde{\mathbf{w}}_t^{\top} \mathbf h_{\rm c} +n_t,\quad t   = 1,\ldots,T,
\end{align}
where $\tilde{\mathbf{w}}_t$ is reflection coefficients at the RIS with elements satisfying the unit modulus constraint during the uplink channel sensing phase,
and $n_t\sim\mathcal{CN}(0,\sigma^2)$ is the additive Gaussian noise. 

Unlike prior work on channel estimation for RIS, this paper considers an active
sensing strategy in which $\tilde{\mathbf{w}}_t$ can be adaptively designed in
each time frame based on the observed pilots in all previous time frames, i.e., 
\begin{align}\label{eq:v_opt}
      \tilde{\mathbf{w}}_{t+1} = \tilde{\mathcal{G}}_t(\tilde{y}_{1:t}, \tilde{\mathbf{w}}_{1:t}), \quad t = 0,\ldots,T-1,
\end{align} 
where $\tilde{\mathcal{G}}_t: \mathbb{C}^{t}\times \mathbb{C}^{tN_r} \rightarrow \mathbb{C}^{N_r}$ is the adaptive sensing scheme that gives the next sensing configuration of the RIS. After collecting the observations in $T$ pilot frames, the downlink reflection coefficients $\tilde{\mathbf{v}}$ at the RIS are then designed as:
\begin{equation}
      \tilde{\mathbf{v}} = \tilde{\mathcal{F}}(\tilde{y}_{1:T},\tilde{\mathbf{w}}_{1:T}),
\end{equation} 
where $\tilde{\mathcal{F}}: \mathbb{C}^{T} \times \mathbb{C}^{TN_r} \rightarrow \mathbb{C}^{N_r}$ is the downlink reflection alignment scheme. Thus the overall problem can be formulated as
\begin{subequations}
      \label{eq:problem_formulation_irs}
      \begin{align}
      \Maximize_{\left\{\tilde{\mathcal{G}}_t(\cdot,\cdot)\right\}_{t=0}^{T-1},\hspace{1pt} \tilde{\mathcal{F}}(\cdot,\cdot) }& \mathbb{E}\left[ \tilde{\mathcal{J}}(\tilde{\mathbf{v}}, \mathbf{h}_{\rm c}) \right]\\
      \text{subject to}\hspace{15pt} &\tilde{\mathbf{w}}_{t+1} = \tilde{\mathcal{G}}_t(\tilde{y}_{1:t}, \tilde{\mathbf{w}}_{1:t}), ~ t = 0,\ldots,T-1,\\
      & \tilde{\mathbf{v}} = \tilde{\mathcal{F}}(\tilde{y}_{1:T},\tilde{\mathbf{w}}_{1:T}).
      \end{align}
\end{subequations}
It is easy to see that the adaptive RIS sensing problem for reflection alignment in \eqref{eq:problem_formulation_irs} has the same form as the generic active learning problem \eqref{eq:problem_formulation_unsup}. Therefore, problem  \eqref{eq:problem_formulation_irs} can be tackled  with the proposed deep active learning framework. 

\subsection{Implementation Details}

To use the existing deep learning libraries that only support real-value operations, we set $\by_t:= [\Re(\tilde{y}_t), \Im(\tilde{y}_t)]^\top$ as the input vector to
the deep active learning unit, $\bw_t :=
\left[\Re\left(\tilde{\bw}_t^\top\right),\Im\left(\tilde{\bw}_t^\top\right)\right]^\top$
as the output of the DNN in the active learning unit, and $\bv :=
\left[\Re\left(\tilde{\bv}^\top\right),\Im\left(\tilde{\bv}^\top\right)\right]^\top$
as the final \changeW{reflective coefficients} designed by the proposed framework. To ensure that
the RIS reflection coefficients in the sensing phase and the downlink data
transmission phase satisfy the unit modulus constraint, we use an
activation function similar to \eqref{eq:activation_hyb} for $\sigma_L^{(\text{w})}(\cdot)$
and $\sigma_R^{(\text{v})}(\cdot)$. Further, we consider an LSTM
cell with $S=512$ states in the deep active learning unit and $4$-layer DNNs
with widths $[1024,1024,1024,M]$ where $M:=2N_r$ in the active learning unit as
well as at the final stage to map the final cell state to the design of the
downlink reflection coefficients. The rest of the implementation details follow
that of the initial beam alignment problem in Section~\ref{sec:imp_det_hybrid}.

\subsection{Numerical Results}

We now show that an adaptive channel sensing approach can improve the performance of the reflection alignment problem in an RIS-assisted system via numerical simulations. The propagation environment between transmitter/receiver and the RIS is modeled by a Rician fading channel with Rician factor $\varepsilon=10$,
for which the channel vector between the transmitter and the RIS, $\mathbf h_{\rm t}$, is given by:
\begin{equation}\label{eq:rician_channel}
    \mathbf h_{\rm t} = \sqrt{\varepsilon \over {1+\varepsilon}} \tilde{\mathbf h}_{\rm t}^{\rm LOS}+ \sqrt{1 \over {1+\varepsilon}}\tilde{\mathbf h}_{\rm t}^{\rm NLOS},
\end{equation}
where $\tilde{\mathbf h}_{\rm t}^{\rm LOS}$ is the line-of-sight component of the channel and $\tilde{\mathbf h}_{\rm t}^{\rm NLOS}  \sim\mathcal{CN}(\mathbf{0},\mathbf{I})$ is the non-line-of-sight component. The line-of-sight component of the channel $\tilde{\mathbf h}^{\rm LOS}_{\rm t}$ is modeled as:
\begin{equation}
      \tilde{\mathbf h}_{\rm t}^{\rm LOS} = \tilde{\alpha}\hspace{1pt}\tilde{\mathbf a}(\theta_{\rm t},\phi_{\rm t}),
\end{equation}
where $\tilde{\alpha}\sim\mathcal{CN}(0,1)$ is the fading coefficient, and $\theta$ and $\phi_{\rm t}$ are respectively the azimuth and elevation angles of arrival from the transmitter to the RIS. Further, we consider a uniform rectangular array RIS with $N_1$ elements per row (horizontal direction) and $N_2$ elements per column (vertical direction), for which the $\ell$th element of the RIS array response vector can be written as \cite{bjornson2020rayleigh}:
\begin{align}
    [\tilde{\mathbf a}(\theta_{\rm t},\phi_{\rm t})]_\ell = e^{j\tfrac{2\pi}{\lambda}\left[i_1(\ell)d_1\sin(\theta_{\rm t})\cos(\phi_{\rm t})+i_2(\ell)d_2\sin(\phi_{\rm t})\right]},
\end{align}
where $d_1$ and $d_2$ are the horizontal and vertical spacings, and $i_1(\ell) = \operatorname{mod}(\ell-1, N_1) $ and $i_2(\ell) =\lfloor (\ell-1)/N_2 \rfloor $ denote the horizontal and vertical indices of element $\ell$, respectively. The channel vector $\mathbf{h}_{\rm r}$ between the receiver and the RIS is modeled in a similar fashion. 

To illustrate the performance of the proposed approach, we consider the following benchmarks.

\textit{Phase matching with perfect CSI:} Assuming that the perfect CSI $\bh_{\rm c}$ is available at the transmitter, the optimal setting of the reflection coefficients is:
\begin{equation}
\tilde{\bv}^\star = [e^{-j\angle h_1^{(\rm c)}  },\cdots, e^{-j\angle h_{N_r}^{(\rm c)}}]^\top,
\end{equation}
where $h_i^{(\rm c)}$ is the $i$th element of the cascaded channel $\bh_{\rm c}$. 
The performance of this optimal design under the perfect CSI assumption serves as the upper bound for the other methods.

\textit{Phase matching with LMMSE CSI estimation:} In this baseline, we first estimate the cascaded channel coefficients by applying linear minimum mean square estimation (LMMSE) to the baseband received signals $\{\tilde{y}_t\}_{t=1}^T$, with the phases of the RIS reflection coefficients set randomly in the sensing phase. The RIS coefficients in the transmission phase are set by matching the phases of the estimated channel. 

\textit{DNN-based design with fixed sensing vector:} 
In this approach, the uplink reflection coefficients $\tilde{\mathbf{w}}_t$ are
again fixed beforehand. 
But instead of estimating the channel, we
design a \changeW{fully connected} DNN to map the received pilots in $T$ time frames,
i.e., $\{\tilde{y}_t\}_{t=1}^T$, to the downlink reflection coefficients
$\tilde{\mathbf v}$ for beamforming gain maximization. 
In the simulations, we use a $4$-layer DNN with widths $[1024,1024,1024,2N_r]$. 

We consider two variants of the above scheme in order to evaluate the benefit of adapting the sensing vector to the channel statistics: 
\begin{itemize}
\item[(i)] The reflection coefficients (i.e., the sensing vectors) are set randomly,
i.e., the phases of $\tilde{\mathbf{w}}_t, \forall t,$ are drawn from a uniform
distribution over $[0,2\pi)$. 
This scheme has been previously proposed in \cite{9427148}. 
\item[(ii)] The reflection coefficients are learned using the deep learning framework.
This is a new scheme that has not appeared in the existing literature to the best of
authors' knowledge. It shows the benefit of intelligently setting the RIS
coefficients in the channel estimation stage. 
\end{itemize}
Finally, we also evaluate the performance of the \textit{DNN-based design with active sensing vector}, based on the proposed active sensing approach using RNN with LSTM.

In the simulations, we consider an $8\times8$ rectangular RIS. The elevation  AoA/AoD between the RIS and the transmitter/receiver is generated uniformly and randomly from $[-\pi/2,\pi/2]$. The azimuth AoA/AoD between the RIS and the transmitter/receiver is generated uniformly and randomly from $[-\pi/2,0]$ and $[0,\pi/2]$, respectively. The SNR of pilot transmission is $0$dB.

\begin{figure}[t]
     \centering
     \includegraphics[width=0.5\textwidth]{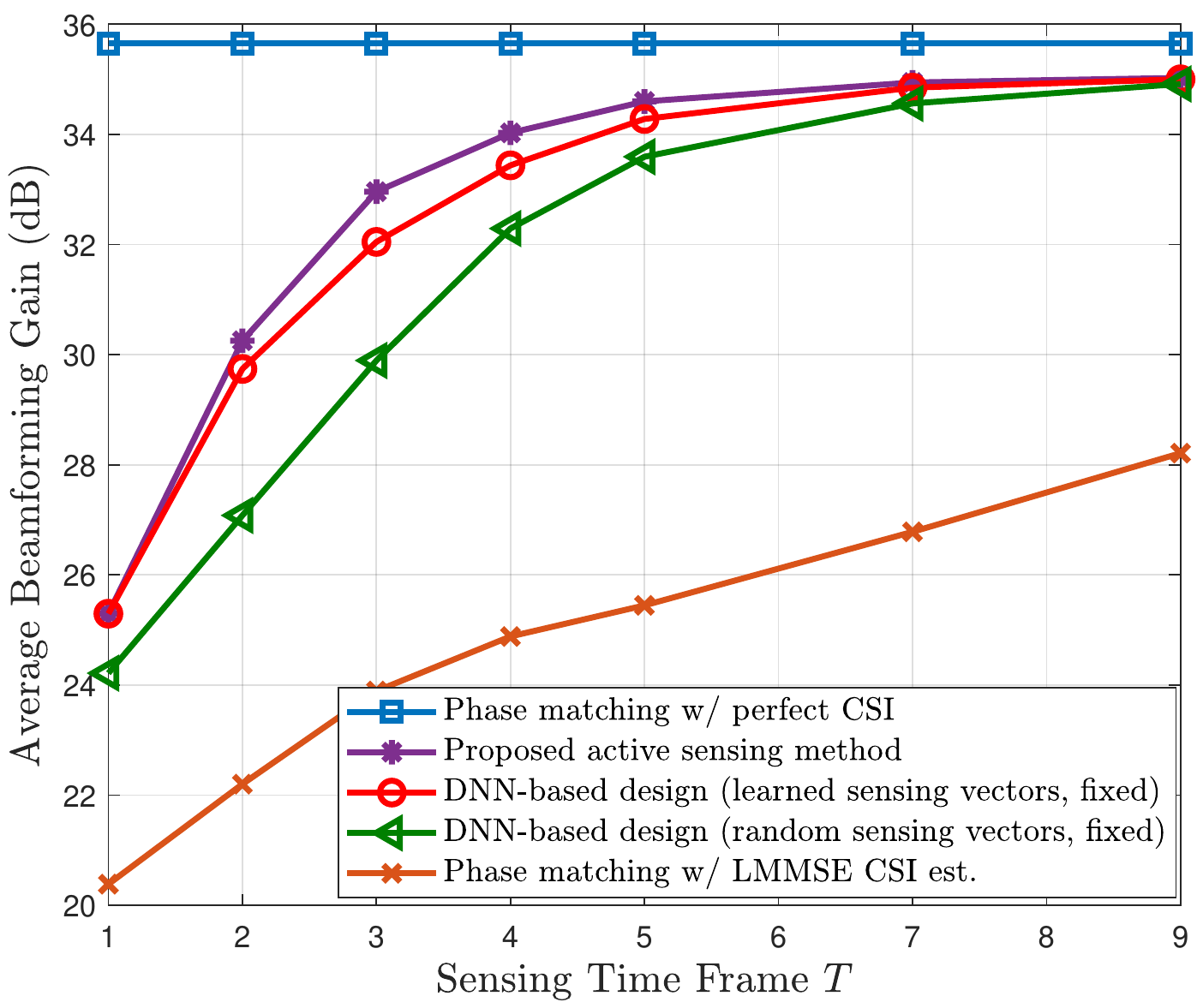}
     \caption{Average beamforming gain in dB versus sensing time frames $T$ for different methods in a system with $N_r = 64$, $\varepsilon=10$, and $\operatorname{SNR}=0$dB.}
     \label{fig:sim_irs}
\end{figure}

In Fig.~\ref{fig:sim_irs}, we plot the average beamforming gain against the
pilot transmission frame length $T$ in the uplink channel estimation phase. It can be
seen that while the achieved beamforming gains for all methods improve as
the number of sensing stages increases, all three \changeW{deep learning based} methods
can achieve much better performance as compared to the traditional approach of
estimating the channels then precoding. This gain is mostly due to 
bypassing the channel estimation step and using a data-driven approach for the
direct maximization of beamforming gain.

Fig.~\ref{fig:sim_irs} further shows that by learning the uplink reflection
coefficients in the DNN-based design from the channel realizations in the
training set, we can further improve the method proposed in \cite{9427148}
which utilizes the random uplink reflection coefficients. Finally, the best
performance is achieved by the active learning framework proposed in this paper.

The results of this paper illustrate that it is important to intelligently adjust 
the reflective coefficients in the channel estimation process of an RIS system. 
Adapting to the channel statistics already provide substantial gain. Active
sensing produces a further benefit. At $T=3$, the downlink performance improvements 
of these two new schemes proposed in this paper as compared to setting the
reflective coefficients randomly in the uplink channel estimation stage as in 
\cite{9427148} are 2dB and an additional 1dB, respectively.
 
\section{Conclusion}
\label{sec:conc} 

This paper develops a deep learning framework for an active sensing setting
where an agent sequentially queries an environment using an adaptive sensing
strategy over multiple time frames for the purpose of maximizing a system
utility. The proposed framework uses an RNN with LSTM units to abstract the
observations obtained so far in each stage into state variables, which are then
used by a fully connected DNN to design the next sensing vector. At the end of
the sensing stages, the final state is mapped
to the desired solution using another DNN.  This paper shows that the proposed
deep active learning framework achieves excellent performance for various
wireless communication applications. In particular, numerical results indicate
that the proposed deep active sensing framework can outperform existing channel
estimation approaches for mmWave beam alignment in an RF-chain-limited massive
\changeF{antenna array} system and for \changeW{reflection} alignment in an RIS-assisted network. 
\changeW{The proposed deep active learning framework has applications beyond the wireless communications setting. As possible future direction, it would be interesting to investigate more general active learning problems.} We also note that the analytic performance characterization for active sensing remains an open research problem.

\bibliographystyle{IEEEtran}
\bibliography{IEEEabrv,referenceF}

\end{document}